\documentclass[fleqn,usenatbib]{mnras}

\usepackage{newtxtext,newtxmath}

\usepackage[T1]{fontenc}

\DeclareRobustCommand{\VAN}[3]{#2}
\let\VANthebibliography\thebibliography
\def\thebibliography{\DeclareRobustCommand{\VAN}[3]{##3}\VANthebibliography}

\usepackage{graphicx}	
\usepackage{amsmath}	
\usepackage{lineno}

\title[Short title, max. 45 characters]{Radiation-Driven Evolution and Gas Kinematics of the Bright-Rimmed Cloud SFO 25}

\author[Porel et al.]{Puja Porel,$^{1, 2}$\thanks{E-mail: pujaporel11@gmail.com}
Archana Soam,$^{1}$
William D. Vacca$^{3}$
and Janik Karoly$^{4}$
\\
$^{1}$Indian Institute of Astrophysics, II Block, Koramangala, Bengaluru 560034, India\\
$^{2}$Pondicherry University, R.V. Nagar, Kalapet, 605014, Puducherry, India\\
$^{3}$NSF’s NOIRLab, 950 N. Cherry Avenue, Tucson, AZ 85719, USA\\
$^{4}$Department of Physics and Astronomy, University College London, Gower Street, London WC1E 6BT, UK
}

\date{Accepted XXX. Received YYY; in original form ZZZ}

\pubyear{\the\year{}}

\begin{document}
\label{firstpage}
\pagerange{\pageref{firstpage}--\pageref{lastpage}}
\maketitle

\begin{abstract}
Bright-rimmed clouds (BRCs) are valuable laboratories for investigating how ionizing radiation from massive stars reshapes molecular clouds and influences star formation. We present a kinematic and dynamical study of the BRC SFO~25 using archival JCMT--HARP observations of the $^{12}$CO, $^{13}$CO, and C$^{18}$O ($J=3\rightarrow2$) transitions at an angular resolution of $\sim14$--15\arcsec\ (0.051--0.054 pc). Gaia parallaxes of young stellar objects associated with the cloud yield a revised distance placing SFO~25 behind the ionizing O7V star HD~47839, implying that ultraviolet radiation can directly affect both the head and tail. The molecular gas exhibits a pronounced head--tail velocity gradient, with the tail systematically redshifted relative to the head, inconsistent with the expectations of classical radiation-driven implosion (RDI) and suggestive of an evolved phase dominated by radiative dispersal and photoevaporation. The head is moderately denser than the tail, while a compact C$^{18}$O clump associated with IRAS~06382+1017 reaches densities of $\sim10^{4},\mathrm{cm^{-3}}$. Virial and energy analyses indicate that the head, tail, and dense clump are gravitationally unbound, with kinetic energy exceeding both gravitational binding energy and external ionized gas pressure. Class~II young stellar object candidates identified in the tail demonstrate that star formation is not confined to the dense head. Although RDI may have triggered earlier star formation in the head, it cannot readily explain the activity observed in the tail, pointing to a more complex evolutionary history than predicted by the classical RDI scenario.
\end{abstract}




\section{Introduction}
\label{section: introduction}

\begin{figure*}
\begin{center}
\resizebox{12.0cm}{12.0cm}{\includegraphics{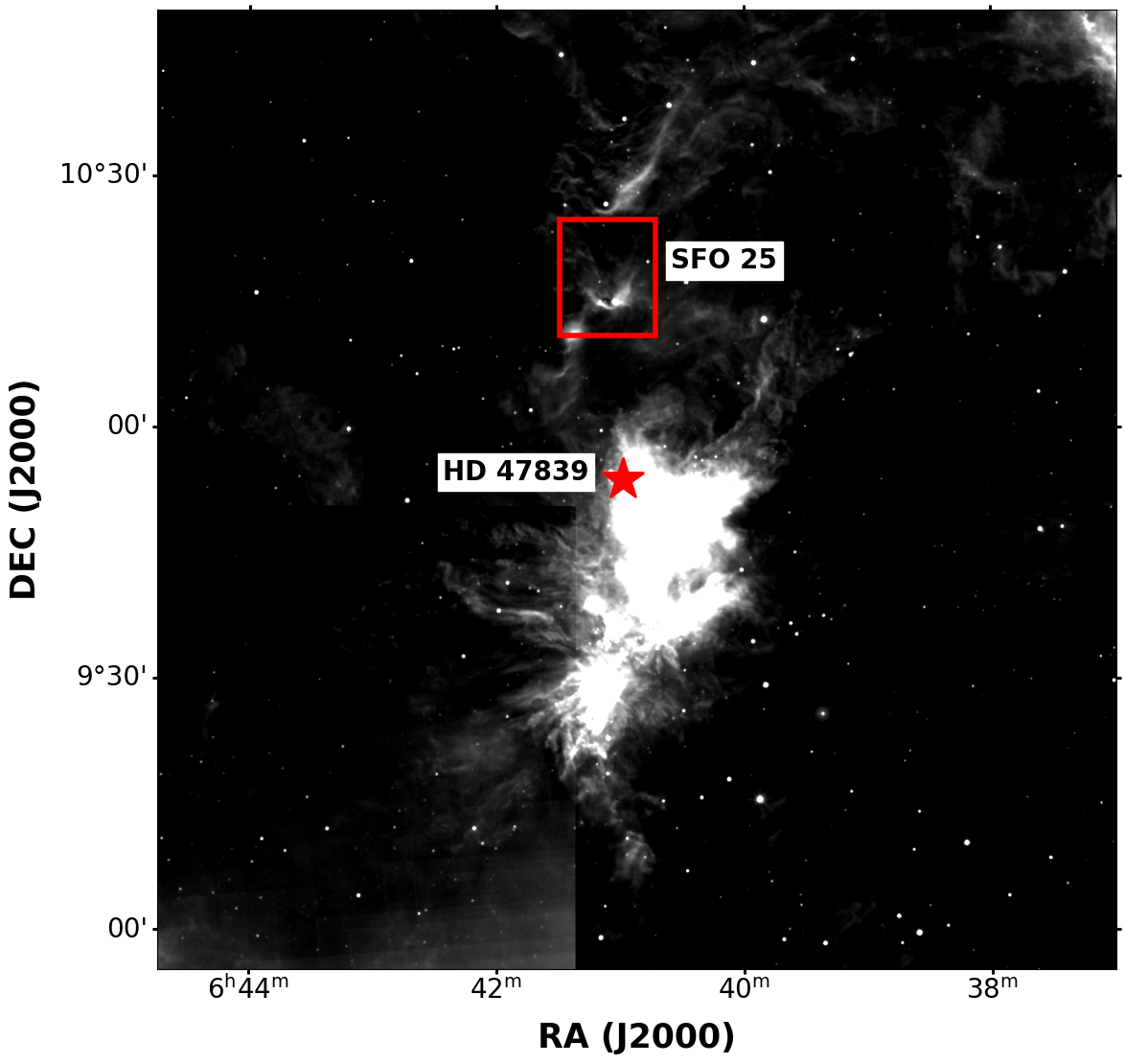}}
\caption{WISE 12\,$\mu$m intensity map covering a field of approximately $2^\circ \times 2^\circ$, shown at an angular resolution of $\sim6.5''$. The red star marks the O7V-type star HD~47839, the primary ionizing source thought to drive the radiation feedback influencing SFO~25. The red box denotes the region mapped in our JCMT--HARP observations, encompassing an area of $\sim11' \times 14'$ around the bright-rimmed cloud SFO 25.
}\label{Fig: wise image}
\end{center}
\end{figure*}

Star formation within dense molecular clouds influenced by nearby massive stars is often driven by external forces, a process known as triggered star formation \citep{elmegreen2011triggered, lee2007triggered, sugitani1991catalog}. One of the most widely discussed mechanisms for externally triggered star formation is Radiation-Driven Implosion (RDI) \citep{sugitani1991catalog}. In this scenario, a pre-existing molecular cloud is exposed to intense far-ultraviolet (FUV) and extreme-ultraviolet (EUV) radiation emitted by nearby massive O- or B-type stars. The EUV photons ionize the surface layers of the cloud, producing a photoionized envelope and initiating the formation of an expanding H\,\textsc{ii} region. At the same time, FUV photons heat the neutral gas in the photodissociation region (PDR), increasing the thermal pressure at the cloud boundary.

The resulting pressure imbalance between the hot, low-density ionized gas and the cold, dense molecular material drives an ionization front into the cloud. This front is typically preceded by a shock front that propagates into the neutral interior, compressing the molecular gas and increasing its density. As the shock penetrates deeper into the cloud, the external layers are progressively eroded through photoevaporation, while the interior becomes increasingly compressed into a dense core or clump.

If the external compression is sufficiently strong, the shocked molecular gas may become gravitationally unstable, leading to collapse and subsequent star formation. This process is often characterized by a sequential evolution: an initial cometary morphology develops, with a dense “head” facing the ionizing source and an elongated “tail” extending away due to material ablation and shadowing effects. The evolution is governed by the competition between ionizing radiation pressure, gas thermal pressure, self-gravity, and magnetic or turbulent support within the cloud.

RDI is therefore considered a key mechanism of triggered star formation at the edges of H\,\textsc{ii} regions, capable of both accelerating collapse in pre-existing density enhancements and reshaping the morphology and kinematics of molecular clouds exposed to strong external radiation fields.

A classic example of such environments is bright-rimmed clouds (BRCs)—dense molecular clouds located at the periphery of H\,\textsc{ii} regions, which exhibit bright ionized rims facing the ionizing source \citep{sugitani1991catalog}. These bright rims are visible in optical wavelengths and result from the ionization of the cloud surface. The ionized gas flows outward towards the ionizing star, while the denser molecular gas is propelled in the opposite direction due to momentum conservation—a phenomenon termed the rocket effect \citep{stahler2008formation, saha2022investigation}. The photoevaporation and subsequent pressure-driven implosion/compression of the cloud interior can lead to the formation of dense condensations (clumps or cores), which are ideal sites for subsequent star formation \citep{bertoldi1989photoevaporation}. Moreover, the asymmetric compression often gives rise to a cometary morphology, characterized by a dense head pointing toward the ionizing source and a more diffuse tail extending away from it \citep{reipurth1983star}.

In this work, we focus on the bright-rimmed cloud SFO 25 \citep{sugitani1991catalog}, situated within the H\,\textsc{ii} region NGC 2264 \citep{sugitani1991catalog, de2002star, morgan2004radio}. The H\,\textsc{ii} region is powered by the O7V-type star HD~47839 \citep{morgan2004radio}, located at a projected distance of approximately 4.7~pc from the bright rim of SFO~25 \citep{morgan2004radio}. A comprehensive discussion of the physical relationship between SFO~25 and HD~47839 is presented in Section~\ref{section: Distance to SFO 25 and Its Physical Association with HD 47839}. Figure \ref{Fig: wise image} presents the location of SFO 25 in WISE 12 $\mu$m emission, where the BRC is enclosed within the red box and the red star symbol marks the position of the ionizing star. Multiple young stellar objects (YSOs) have been identified within SFO~25 in various studies \citep{sugitani1991catalog, hosoya2020spectroscopic, morgan2008scuba, wolf2003star, buckle2015wide}, indicating recent star formation in this region. One of these YSOs corresponds to the infrared source IRAS~06382+1017 \citep{sugitani1991catalog}, believed to be a Class~I protostar exhibiting molecular outflow activity \citep{wolf2003star}. The IRAS source is also associated with several Herbig--Haro (HH) objects—such as HH~124A--F, HH~273A, and HH~274—and radio sources (e.g., HH~124~VLA~1,~2,~7--10) identified in previous studies \citep{walsh1992two, reipurth2002radio, reipurth2004deep}. The kinematics of the molecular outflow from this source were examined by \cite{de2002star}.

Our analysis is based on archival JCMT--HARP observations of SFO~25 in the $^{12}$CO, $^{13}$CO, and C$^{18}$O $J = 3 \rightarrow 2$ transitions, which are used to
investigate the physical conditions and dynamical state of the cloud. We examine the head and tail components of SFO~25 to evaluate their star-forming potential by deriving
key physical parameters, including molecular mass, virial stability, and energy budget. In addition, we identify a prominent C$^{18}$O clump and
analyze its internal energy budget to determine whether it is gravitationally bound and
capable of future star formation, or instead undergoing radiative disruption.

Using the ALLWISE catalog, we identify two Class~II protostars located within the tail region of SFO~25 for the first time. A detailed description of the identification methodology is provided in Section~\ref{section: Distance to SFO 25 and Its Physical Association with HD 47839}. While radiation-driven implosion was the plausible mechanism for triggering star formation in the dense head of SFO~25 in the early stage of evolution, the physical processes responsible for the star formation activity in the tail region remain highly uncertain and are discussed in detail in Section~\ref{section: Star Formation Status and Potential in SFO 25}.

Our broader goal is to reconstruct the internal structure and kinematic evolution of
SFO~25 under the influence of external ionizing radiation. To this end, we estimate quantities such as the systemic velocity, virial parameter, energy budget, photoevaporative mass-loss rate, and depletion timescale for both the head and tail components—parameters that are crucial for assessing whether the cloud is likely to undergo gravitational collapse or progressive dispersal. While earlier studies primarily focused on star formation activity within the head of SFO~25, without addressing the long-term evolutionary trajectory of
the cloud, our analysis provides a global assessment of the star-forming potential of the entire system and offers constraints on its future evolution.

We have also re-estimated the distance to SFO~25 using young stellar objects physically
associated with the cloud, based on modern Gaia~Data Release (DR3) \citep{vallenari2023gaia} astrometry in combination with infrared identifications from the ALLWISE catalog. Previous studies adopted a distance of $\sim 780$~pc derived from photometric measurements by \citet{turner1976value}; however, the availability of precise Gaia parallaxes motivates a reassessment of this value. Our updated distance determination provides a more robust foundation for deriving
distance-dependent physical parameters and for evaluating the physical association between SFO~25 and its ionizing source.

The structure of this paper is as follows: Section~\ref{section: data} describes the observational datasets used in this study. Section~\ref{section: Results} presents the key results, including morphological and kinematic features. In Section~\ref{section: analysis}, we derive physical parameters to assess star-forming potential. Section~\ref{section: discussion} interprets the findings in the context of RDI and triggered star formation, and finally, Section~\ref{section: conclusion} summarizes our main conclusions.

\section{Archive data} \label{section: data}

We analyzed archival data of the $^{12}$CO, $^{13}$CO, and C$^{18}$O ($J = 3 \rightarrow 2$) rotational transitions obtained with the James Clerk Maxwell Telescope (JCMT), covering an area of $\sim11^\prime \times 14^\prime$ toward the SFO~25 region. These transitions probe moderately dense molecular gas and provide constraints on the kinematics; however, excitation conditions cannot be independently determined from a single transition per isotopologue and require additional modelling assumptions such as LTE or non-LTE radiative transfer. The data were acquired on 2008 October 11 under program M08BU15 using the Heterodyne Array Receiver Programme (HARP) in raster-scan mode with a scan spacing of 29.1 arcsec. The observations were carried out in \texttt{PSSW} (position-switching) mode, where emission-free reference positions were used for sky subtraction. For the $^{12}$CO observations, the atmospheric opacity at 225\,GHz, measured by the Caltech Submillimeter Observatory (CSO) 225\,GHz tau monitor, ranged from $\tau_{225} = 0.094$ at the start to $\tau_{225} = 0.088$ at the end, corresponding to JCMT weather band~3. For the $^{13}$CO and C$^{18}$O observations, the atmospheric opacity was lower, ranging from $\tau_{225} = 0.062$ at the start to $\tau_{225} = 0.056$ at the end, corresponding to JCMT weather band~2. The median system temperatures were 296, 337, and 407~K for $^{12}$CO, $^{13}$CO, and C$^{18}$O, respectively. The effective on-source integration times were 6~min for $^{12}$CO and 11~min for both $^{13}$CO and C$^{18}$O.

Data reduction was carried out using the ORAC-DR pipeline within the STARLINK software suite, employing the \texttt{REDUCE\_SCIENCE\_NARROWLINE} recipe \citep{jenness2015orac, buckle2010jcmt}. The spectra were corrected for atmospheric attenuation using the JCMT standard calibration model and converted from antenna temperature ($T_A$) to main-beam temperature ($T_{\mathrm{mb}} = T_A / \eta_{\mathrm{mb}}$). A main-beam efficiency of $\eta_{\mathrm{mb}} = 0.61$ was adopted for all transitions \citep{buckle2010jcmt}. Linear baselines were removed, and the data were regridded onto a uniform $7.3\arcsec$ pixel grid in equatorial coordinates. A summary of the observational parameters, including rest frequencies, velocity and angular resolutions, and mean rms noise levels for all transitions, is given in Table~\ref{Table: table co}.

\begin{table*}
\begin{center}
	\caption{Observed CO isotopologues transitions and observational parameters.}
	\label{physical properties of global cloud and clumps}
    \renewcommand{\arraystretch}{1.3} 
	\begin{tabular}{lcccc} 
		\hline
		Transition & Rest frequency & Velocity resolution  & Angular resolution & Mean 1 $\sigma$ rms \\ 
		  & (GHz) & (km s$^{-1}$) & ($\arcsec$) & (K)\\ 
		\hline
    $^{12}$CO (3--2)  & 345.79 & 0.42  & 14 & 0.4 \\
    $^{13}$CO (3--2)  & 330.59 & 0.056 & 15 & 1.1 \\
    C$^{18}$O (3--2)  & 329.33 & 0.056 & 15 & 1.3 \\
		\hline 
	\end{tabular}
    \label{Table: table co}
\end{center}
\end{table*}

\section{Results} \label{section: Results}

\begin{figure*}
\begin{center}
\resizebox{16.5cm}{6.5cm}{\includegraphics{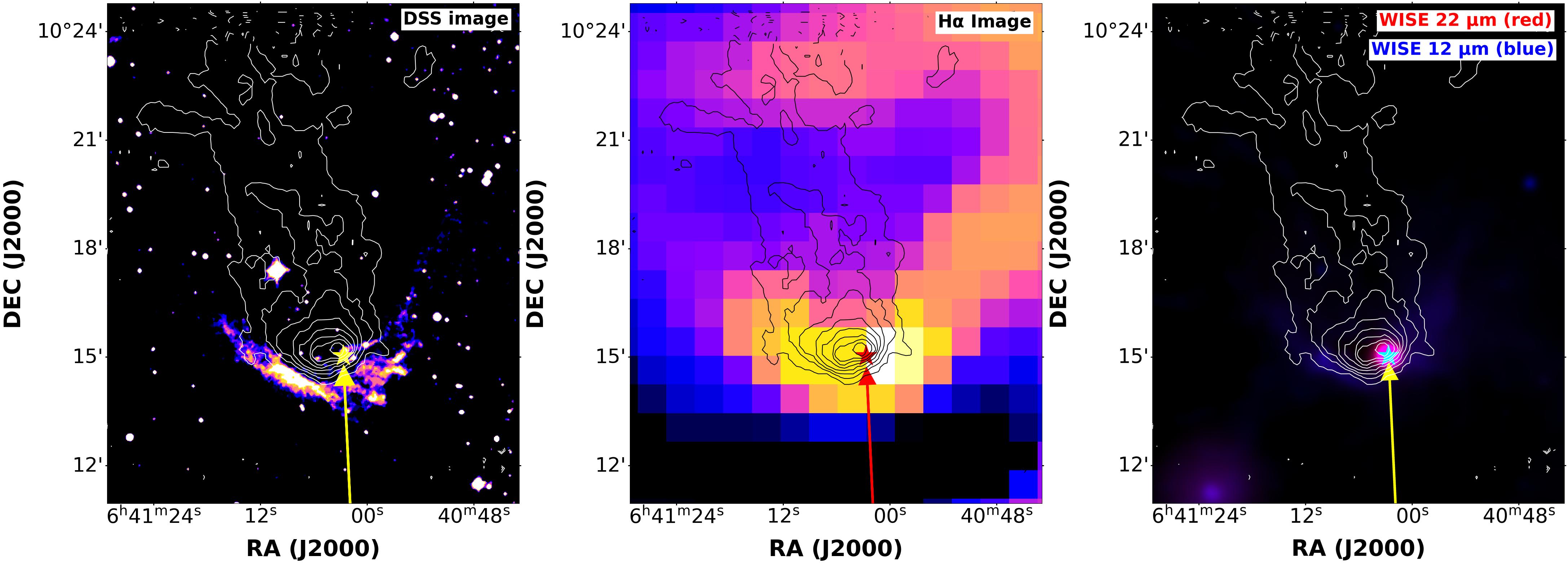}}
\caption{\textbf{Left:} Optical DSS image of the bright-rimmed cloud SFO~25 observed at a wavelength of $\sim$536\,nm, with an angular resolution of $\sim$2$^{\prime\prime}$. \textbf{Middle:} H$\alpha$ emission map of SFO~25 obtained from the SHASSA via NASA SkyView, with an angular resolution of $\sim$3.2$^{\prime}$.  
\textbf{Right:} Two-color composite image of the SFO~25 region constructed from WISE 22\,$\mu$m (red) and 12\,$\mu$m (blue) emissions, corresponding to angular resolutions of $\sim$12$^{\prime\prime}$ and $\sim$6.4$^{\prime\prime}$, respectively.  
Overlaid black and white contours represent the $^{13}$CO (3--2) integrated intensity levels at 3$\sigma$, 6$\sigma$, 10$\sigma$, 15$\sigma$, 20$\sigma$, 25$\sigma$, 30$\sigma$, and 35$\sigma$, where $\sigma$ denotes the rms noise of the background emission.  
The yellow, red, and cyan star symbols indicate the position of IRAS~06382+1017, while the red and yellow arrows mark the projected direction of incident ultraviolet radiation from the ionizing O7V-type star HD~47839. Each panel covers a field of view of approximately $11^{\prime} \times 14^{\prime}$.
}
\label{Fig: wise and optical image}
\end{center}
\end{figure*}

\begin{figure*}
\begin{center}
\resizebox{12.0cm}{13.5cm}{\includegraphics{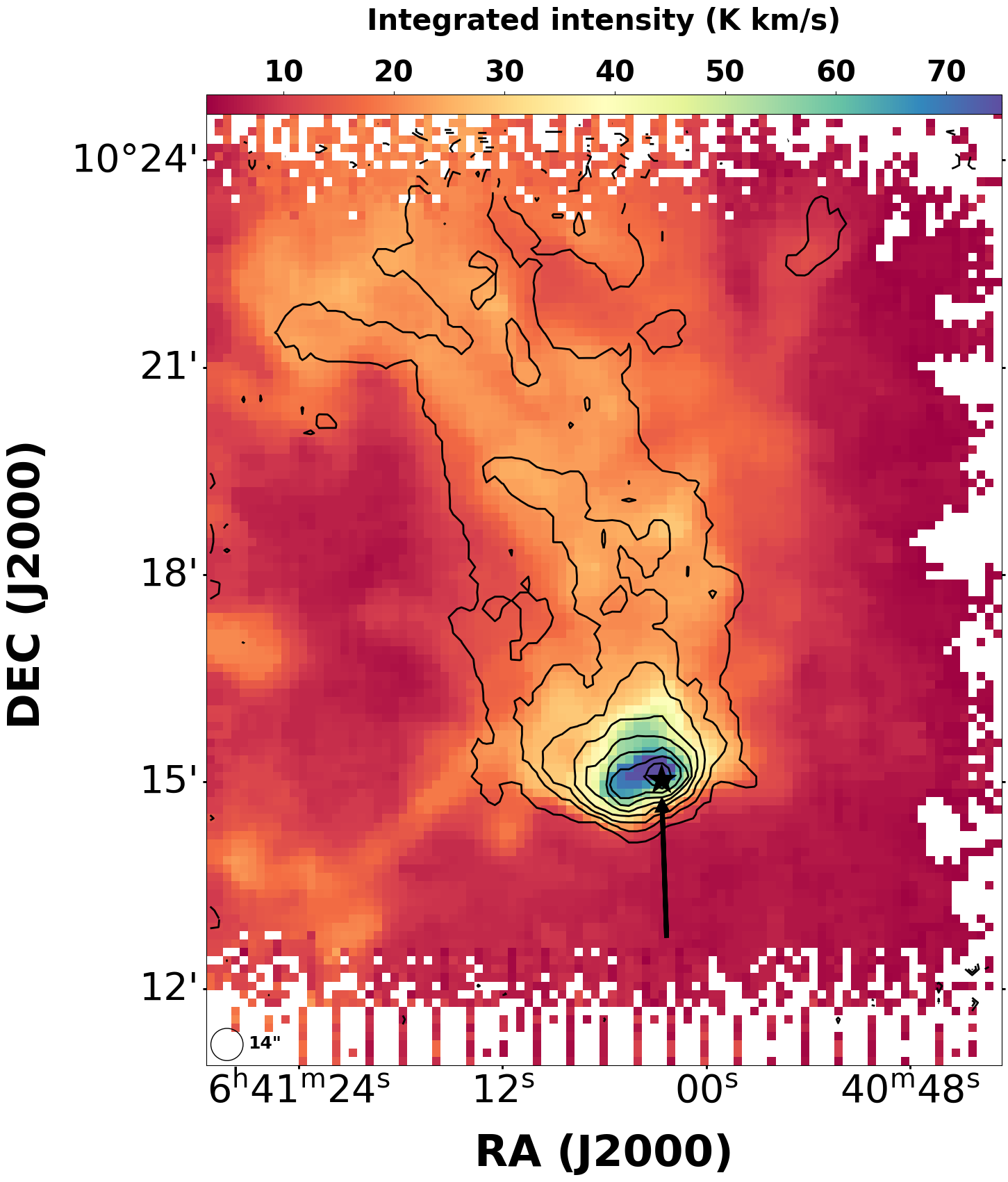}}
\caption{Integrated intensity (moment~0) map of the $^{12}$CO (3--2) emission, constructed over the velocity range from -0.03 to 12.25\,km\,s$^{-1}$ above the 3$\sigma$ threshold, where $\sigma$ ($\approx$\,1.03\,K\,km\,s$^{-1}$) represents the standard deviation of the background noise. Overlaid in black are the integrated intensity contours of the $^{13}$CO (3--2) emission, constructed over the velocity range of 5.35 to 10.06\,km\,s$^{-1}$, beginning at 3$\sigma$ and increasing through 6$\sigma$, 10$\sigma$, 15$\sigma$, 20$\sigma$, 25$\sigma$, 30$\sigma$, and 35$\sigma$, where $\sigma$ denotes the rms background noise level of $\sim$0.56\,K\,km\,s$^{-1}$. The black star indicates the position of the embedded IRAS source IRAS~06382+1017, while the black arrow marks the projected direction of the incident ionizing radiation from the nearby O7V-type star HD~47839. The black circle in the lower-left corner represents the JCMT-HARP beam size for the $^{12}$CO (3--2) observation.}\label{Fig: moment 0 map}
\end{center}
\end{figure*}

\begin{figure*}
\begin{center}
\resizebox{16.0cm}{8.5cm}{\includegraphics{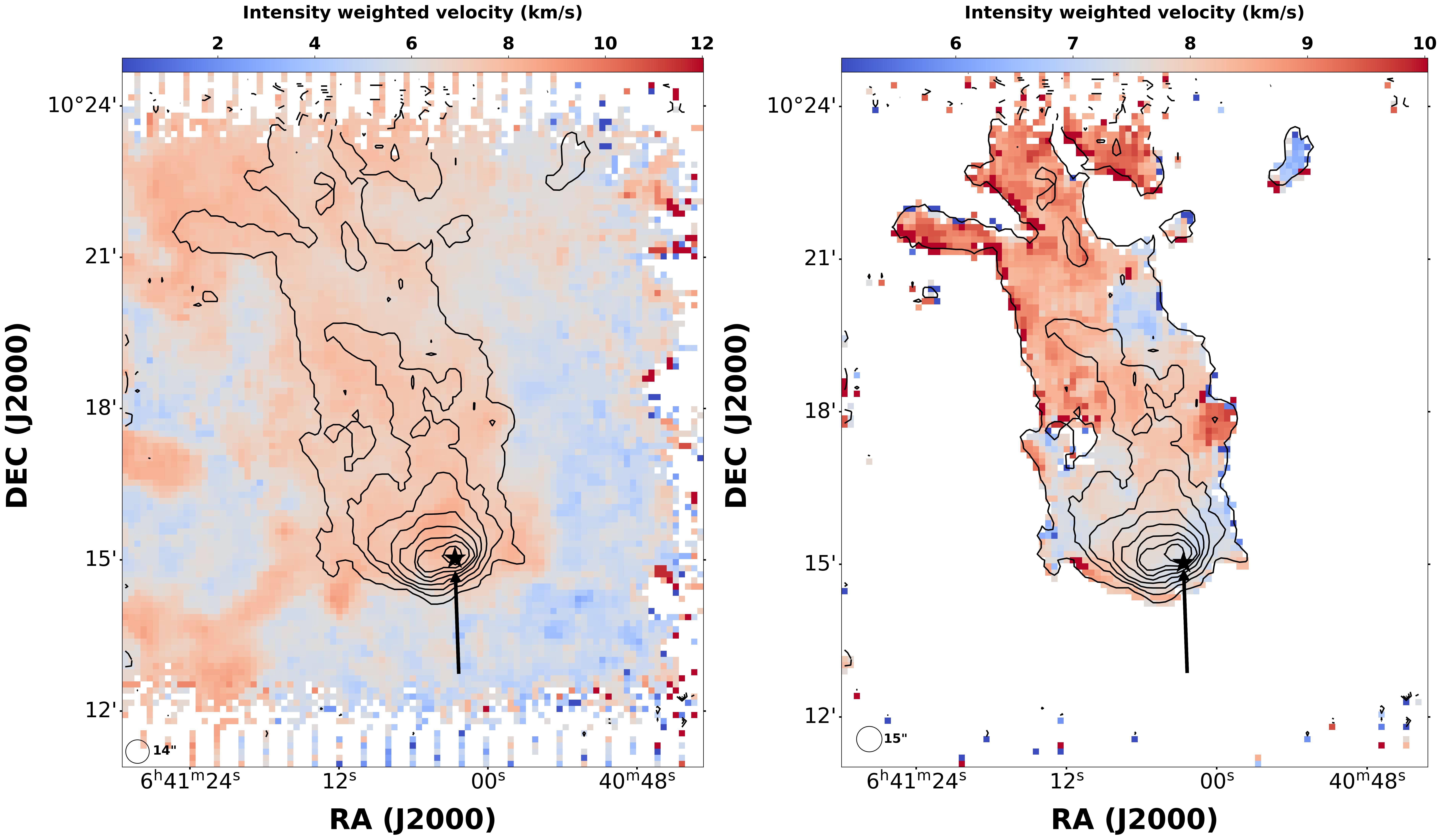}}
\caption{Intensity-weighted velocity (moment~1) maps of the $^{12}$CO (3--2) emission (left) and $^{13}$CO (3--2) emission (right), integrated over the same velocity ranges as their respective moment~0 maps. 
Both maps are overlaid with the $^{13}$CO (3--2) integrated intensity contours (in black), corresponding to the same contour levels as those shown in Figure~\ref{Fig: moment 0 map}. 
The black star indicates the position of the embedded infrared source IRAS~06382+1017, while the black arrow shows the projected direction of the incident ionizing radiation from the nearby O star HD~47839.}\label{Fig: moment 1 map}
\end{center}
\end{figure*}

\begin{figure*}
\begin{center}
\resizebox{16.0cm}{12.5cm}{\includegraphics{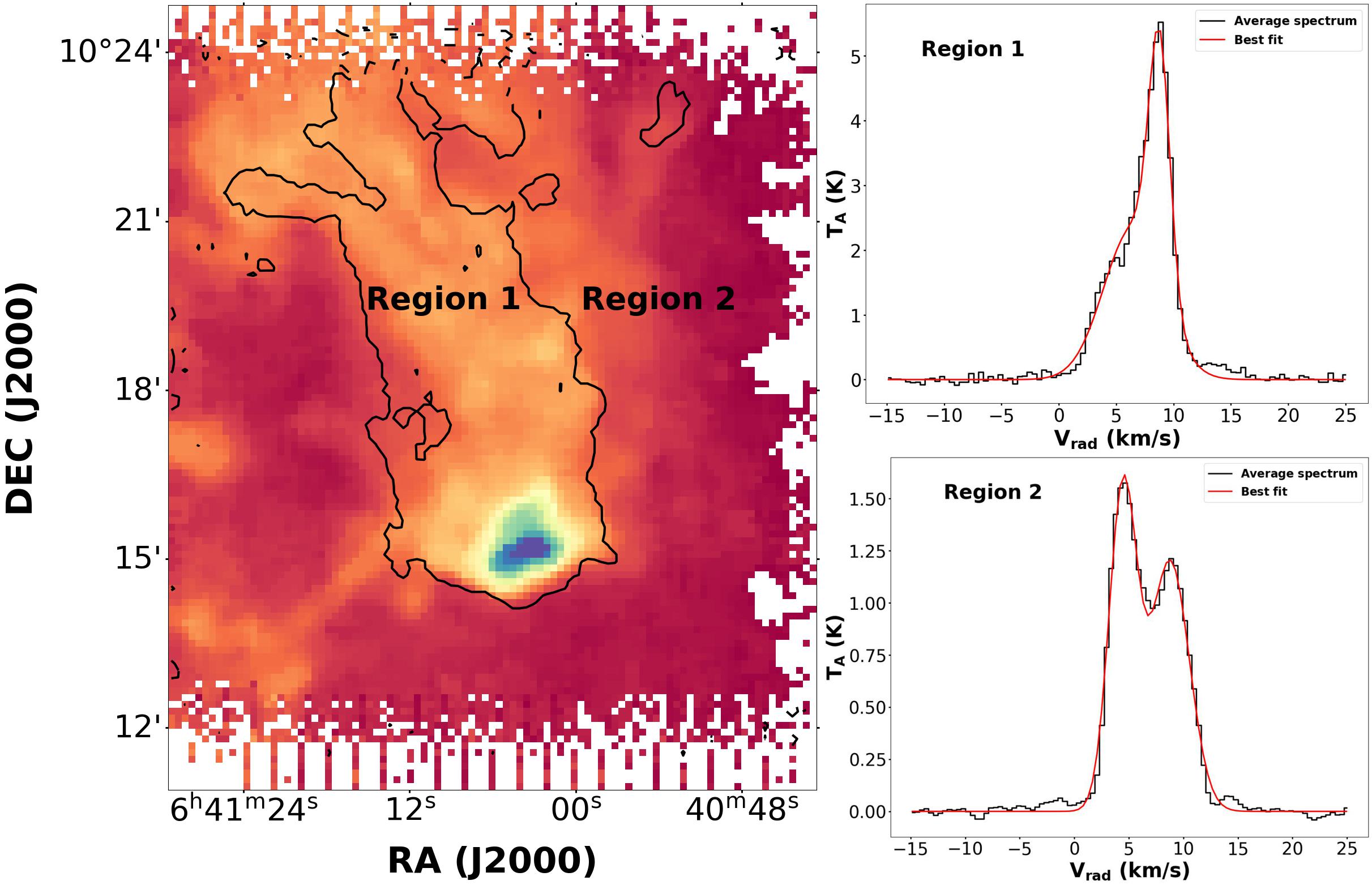}}
\caption{\textbf{Left:} The spatial segmentation of two distinct regions—Region 1, and Region 2. Overlaid on the map is the outermost contour of the $^{13}$CO (3--2) integrated emission, shown in black.
\textbf{Right:} Spectral profiles of the $^{12}$CO (3--2) emission corresponding to Region 1 (top), and Region 2 (bottom) are shown in black, with red curves indicating the best-fit Gaussian models.}\label{Fig: Regions in 12co}
\end{center}
\end{figure*}

\begin{figure*}
\begin{center}
\resizebox{11.0cm}{9.5cm}{\includegraphics{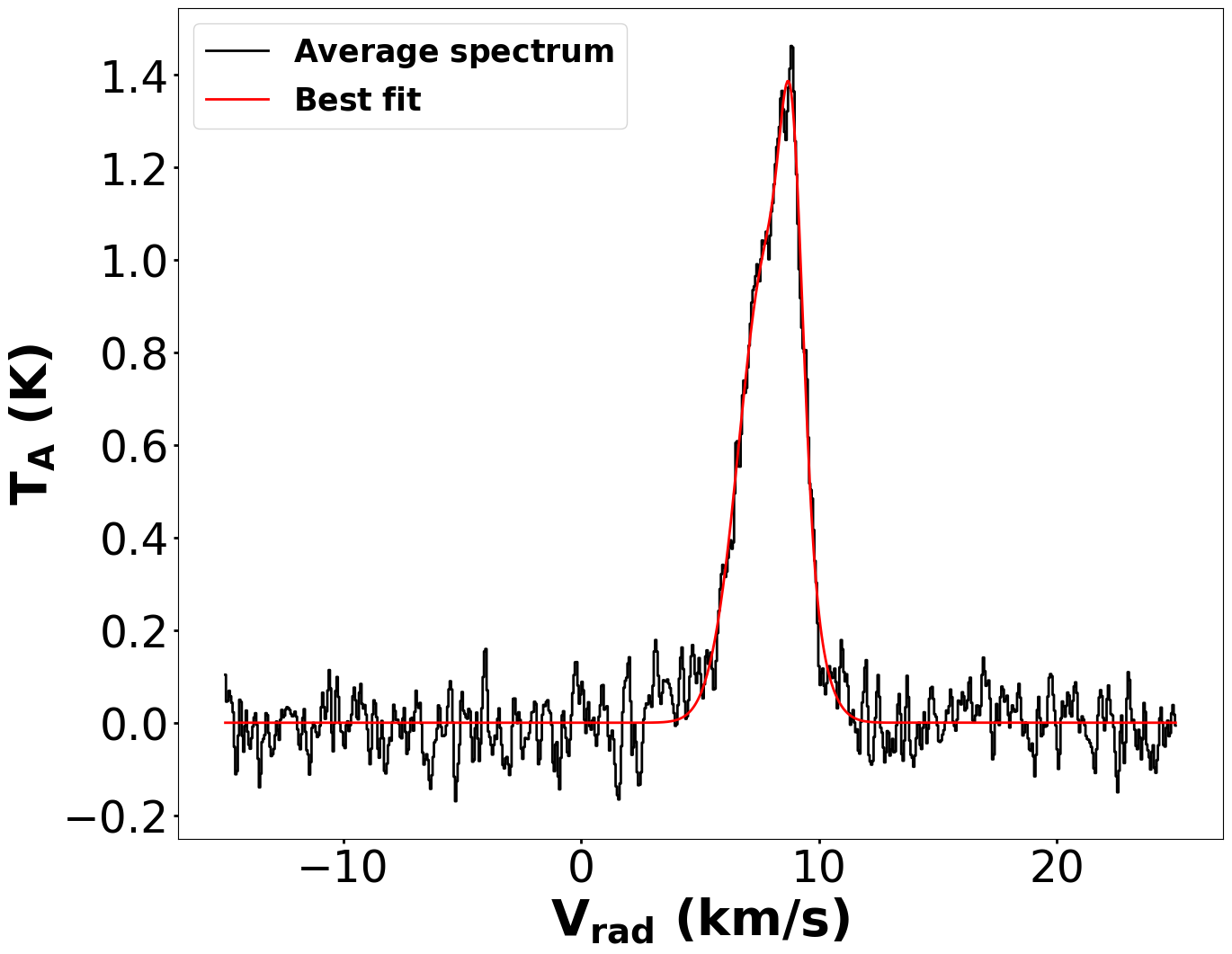}}
\caption{Average spectral profile of the entire $^{13}$CO (3--2) emission region extracted from $^{13}$CO (3--2) data, shown in black. The red curve represents the corresponding best-fit Gaussian model to the $^{13}$CO spectrum.
}\label{Fig: sfo spectrum}
\end{center}
\end{figure*}

\begin{figure*}
\begin{center}
\resizebox{15.0cm}{9.5cm}{\includegraphics{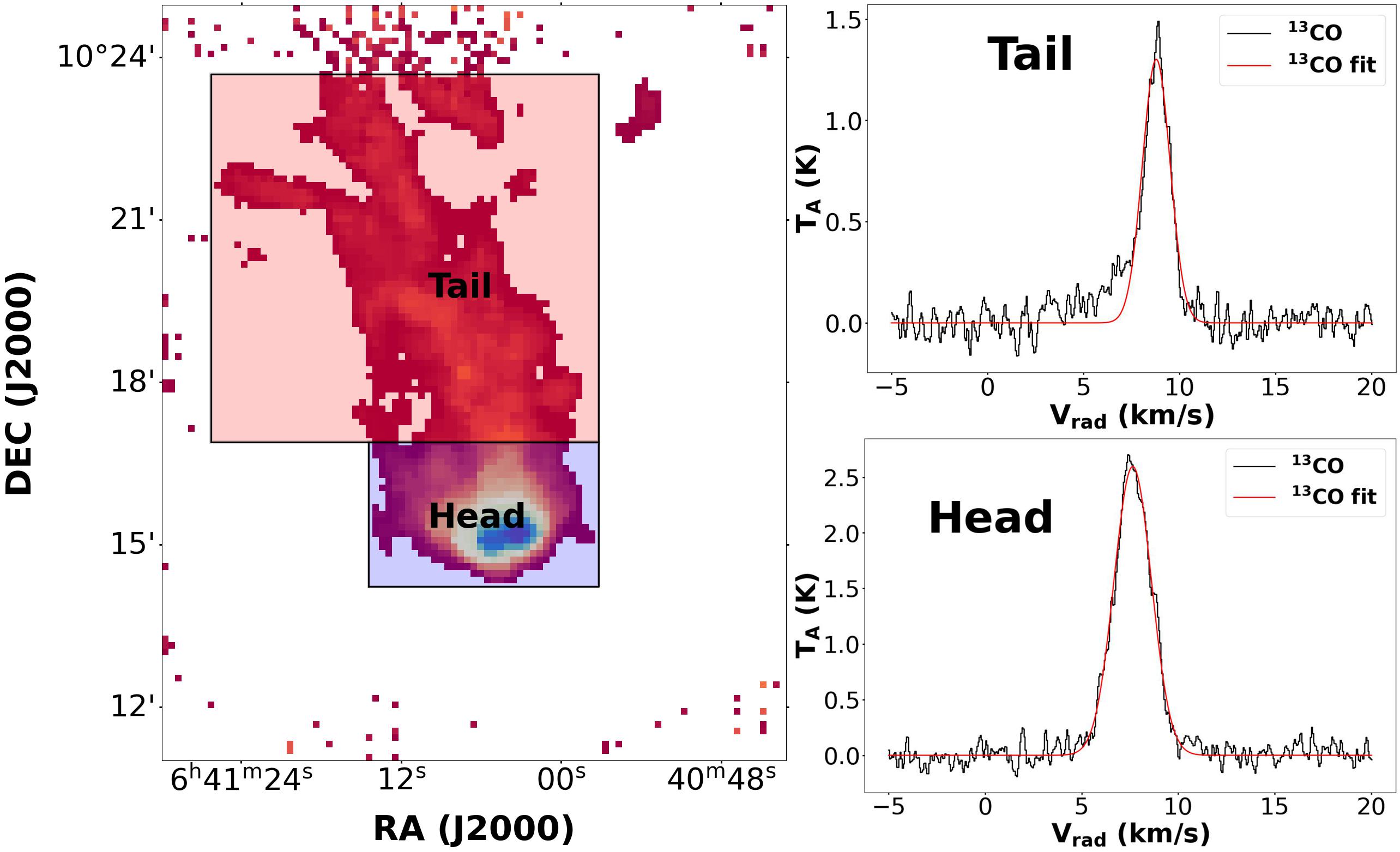}}
\caption{\textbf{Right:} Schematic depiction highlighting the head and tail structures of the SFO~25. \textbf{Left:} $^{13}$CO (3--2) spectral profiles corresponding to the tail (top panel) and head (bottom panel) regions of SFO~25. The red curves represent Gaussian fits to the observed line profiles.
}\label{Fig: head, tail of 13CO}
\end{center}
\end{figure*}

The left panel of Figure~\ref{Fig: wise and optical image} presents the optical Digitized Sky Survey (DSS) image of the BRC SFO~25  at a wavelength of approximately 536~nm. This optical band primarily traces continuum starlight scattered by dust grains on the illuminated surface of the cloud, thereby delineating the ionization boundary between the ionized and molecular regions. The molecular gas distribution is shown by the white (left and right) and black (middle) contours representing the $^{13}$CO~(3--2) integrated emission. The prominent bright rim, clearly visible ahead of the molecular condensation, marks the surface layer where UV photons from the nearby ionizing star impinge upon and heat the dense molecular material. 
The middle panel displays the H$\alpha$ image of SFO~25, obtained from the Southern H-Alpha Sky Survey Atlas (SHASSA) via NASA SkyView. In contrast to the optical continuum, the H$\alpha$ emission traces the recombination radiation from ionized hydrogen gas, thereby revealing both the bright rim and the partially ionized gas extending into or just behind the molecular head of SFO~25. This morphology indicates that the ionization front has penetrated the outer layers of the cloud, consistent with active photoionization and gradual erosion driven by the UV radiation field within the H\,{\textsc{ii}} region NGC~2264.
The right panel shows a two-color composite image, where WISE 22~$\mu$m emission is displayed in red and WISE 12~$\mu$m emission in blue. The 12~$\mu$m band primarily traces polycyclic aromatic hydrocarbons (PAHs) excited within this bright-rimmed cloud, while the 22~$\mu$m band is dominated by thermal emission from warm dust grains. Both infrared components exhibit strong emission concentrated around the IRAS point source (marked by a cyan star). The concentration of PAH and warm-dust emission near IRAS 06382+1017 reflects strong local heating by the embedded protostar (and associated outflow activity) and does not imply the absence of external illumination of the cloud surface. As different tracers probe different layers, mid-IR bands preferentially highlight locally heated dust and PAH emission, while optical/H$\alpha$ and CO moment maps trace the ionization front and large-scale gas kinematics shaped by the O-star, respectively.

Figure~\ref{Fig: moment 0 map} presents the integrated intensity (moment~0) map of the $^{12}$CO (3--2) emission, integrated over the velocity range of $-0.03$ to $12.25$~km~s$^{-1}$, and displayed above the $3\sigma$ threshold, where $\sigma$ denotes the standard deviation of the background noise. Overlaid in black contours is the integrated intensity distribution of the $^{13}$CO (3--2) emission, integrated over the velocity range of $5.35$ to $10.06$~km~s$^{-1}$ and also plotted above the $3\sigma$ level. These velocity intervals were selected to encompass channels where the signal in both molecular tracers exceeds their respective mean $1\sigma$ rms noise levels, as listed in Table~\ref{Table: table co}. Owing to the optically thick nature of $^{12}$CO and the moderately optically thin nature of $^{13}$CO, the latter serves as a more reliable tracer of the dense molecular gas that delineates the main body of the bright-rimmed cloud SFO~25. The $^{12}$CO emission appears bright within the main body of SFO~25, while becoming diffuse beyond the $^{13}$CO$ $ contours, indicating the presence of extended, low-density molecular gas surrounding the cloud. This inference is consistent with the findings of \citet{de2002star}. The southern portion of the cloud exhibits comparatively stronger emission in both $^{12}$CO and $^{13}$CO lines, suggesting a higher degree of compression relative to the northern region. The presence of an IRAS point source embedded within this dense molecular zone further supports the scenario of enhanced compression and recent star formation in the southern part of SFO~25.

Figure~\ref{Fig: moment 1 map} displays the intensity-weighted velocity (moment~1) maps of the $^{12}$CO (3--2) and $^{13}$CO (3--2) transitions, shown in the left and right panels, respectively. These maps were generated over the same velocity ranges as their corresponding moment~0 maps, with the $^{13}$CO integrated intensity contours overlaid in black for spatial reference. In the $^{12}$CO moment~1 map, the region encompassed by the outermost $^{13}$CO contour (3$\sigma$ level) predominantly exhibits red-shifted velocities, while the surrounding areas outside the main body of SFO~25 display a mixture of red- and blue-shifted components.

Based on the observed velocity structure, the $^{12}$CO emission was separated into two distinct subregions: Region~1, corresponding to the area enclosed by the outermost $^{13}$CO integrated intensity contour, and Region~2, representing the more extended, diffuse emission located outside the main molecular body, as illustrated in the left panel of Figure~\ref{Fig: Regions in 12co}. The upper and lower right panels display the corresponding $^{12}$CO spectra for these two regions, together with their best-fit Gaussian models. Region~1 exhibits two velocity components centered at approximately 5.12~km~s$^{-1}$ and 8.78~km~s$^{-1}$, with the redshifted component being dominant. In contrast, Region~2 also shows two well-defined velocity components near 4.47~km~s$^{-1}$ and 8.78~km~s$^{-1}$, where the blueshifted component is comparatively stronger.

These spectral signatures clearly indicate that the main body of SFO~25 is red-shifted relative to its surroundings, while the peripheral molecular gas is dominated by blue-shifted emission. The appearance of a weak blue component in the Region~1 spectrum, along with a red component in Region~2 similar to the red component of Region 1, suggests partial kinematic coupling or gas mixing between these adjacent layers. Such a velocity pattern is consistent with the dynamical influence of the ionizing radiation from the nearby O-type star HD~47839, whose intense radiation pressure and photoionization-driven flow compress and accelerate the molecular gas away from the observer, thereby producing a red-shifted core which is discussed in more detail in Section~\ref{section: discussion}.

To determine the systemic velocity of SFO~25, we analyzed the average $^{13}$CO~(3--2) spectrum extracted over the entire molecular emission region. Figure~\ref{Fig: sfo spectrum} displays the observed $^{13}$CO spectral profile (in black) along with its best-fit Gaussian components (in red). The spectrum reveals two distinct velocity components centered at approximately 7.86~km~s$^{-1}$ and 8.87~km~s$^{-1}$, with the redshifted component being dominant. 

Given that $^{13}$CO is a reliable tracer of moderately dense molecular gas in SFO~25, the emission was further subdivided into the head and tail substructures. The head region is characterized by a concentration of high-density gas, as indicated by the prominent HCO$^{+}$(3--2) emission coincident with this area. This correspondence aligns with the compact, dense region marked by the black box in Figure~19 of \citet{de2002star}, thereby confirming the physical distinction between the head and tail components, with the head representing the denser lower region and the tail corresponding to the more diffuse upper region representing a cometary morphology of the SFO 25 cloud. Figure~\ref{Fig: head, tail of 13CO} illustrates this classification, where the left panel delineates the spatial extent of the head and tail regions shown in the integrated intensity map of $^{13}$CO (3--2) emission, and the right panel presents their corresponding $^{13}$CO spectral profiles. Gaussian fitting of these spectra yields systemic velocities of 7.65~km~s$^{-1}$ for the head and 8.79~km~s$^{-1}$ for the tail, indicating that the tail is red-shifted with respect to the head. The maximum integrated intensity within the head region coincides with the position of the IRAS point source, suggesting that the source is deeply embedded within a dense molecular core.

\section{Analysis}
\label{section: analysis}

\subsection{Distance to SFO 25 and Its Physical Association with HD 47839}
\label{section: Distance to SFO 25 and Its Physical Association with HD 47839}

\begin{figure*}
\begin{center}
\resizebox{14.0cm}{14.0cm}{\includegraphics{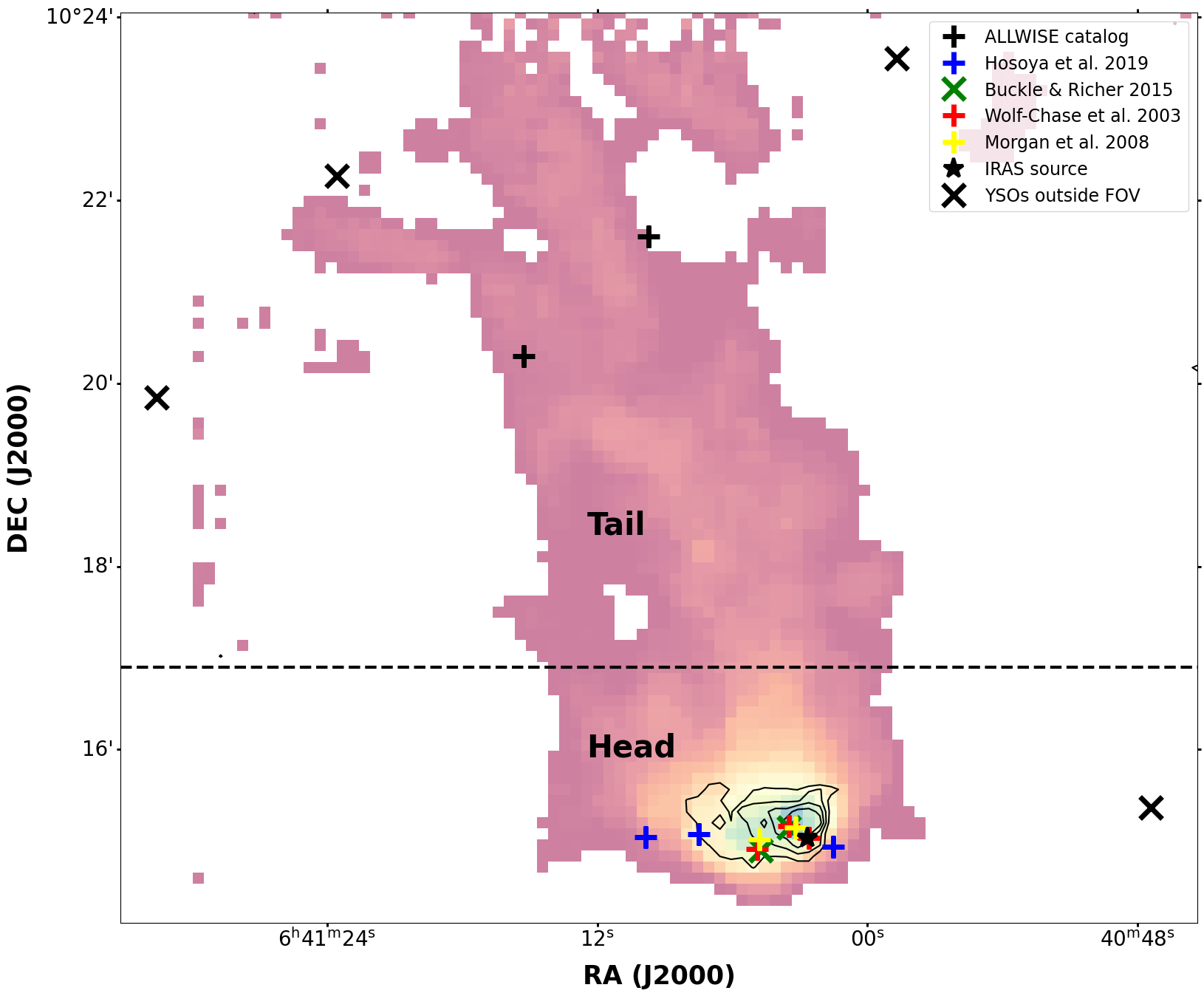}}
\caption{Spatial distribution of young stellar object candidates superposed on the $^{13}\mathrm{CO}$ integrated intensity map. 
The black contours trace the $\mathrm{C}^{18}\mathrm{O}$ integrated emission. 
Blue plus symbols mark the positions of spectroscopically confirmed YSOs from \citet{hosoya2020spectroscopic}, while green crosses indicate YSOs identified by \citet{buckle2015wide}. 
Red plus symbols denote YSOs reported by \citet{wolf2003star}, 
and yellow plus symbols represent YSOs detected by \citet{morgan2008scuba}. 
The black star marks the position of the IRAS source cataloged by \citet{sugitani1991catalog}. 
The black plus and black cross symbols correspond to YSO candidates independently identified in this work using the ALLWISE catalog, 
following the color-based classification criteria of \citet{koenig2011wide}. Among these newly identified sources, only those marked by black plus symbols lie within the observed field of view of our SFO 25 observations. The black dashed line delineates the boundary between the head and tail regions of SFO 25.}
\label{Fig: star formation}
\end{center}
\end{figure*}

Previous studies have adopted a distance of 780~pc for the bright-rimmed cloud SFO~25, based on distance-modulus estimates reported by \citet{turner1976value} and subsequently used by \citet{sugitani1991catalog} and \citet{de2002star}. In contrast, the distance to the O-type star HD~47839, widely regarded as the dominant source of ionizing radiation affecting SFO~25 \citep{sugitani1991catalog, de2002star, morgan2004radio}, is estimated to be $\sim$713~pc from Gaia DR3 astrometry \citep{vallenari2023gaia} satisfying the criteria RUWE $< 1.4$. Since the distances to the BRC and its proposed ionizing source were derived using different methodologies, a re-determination of the distance to SFO~25 using modern astrometric data is warranted.

In this work, we estimate the distance to SFO~25 using Gaia-based distances of young stellar objects physically associated with the cloud. Gaia parallaxes were assigned to YSOs by positional cross-matching infrared-identified sources with the Gaia database within a matching radius of $2\arcsec$. The IRAS source IRAS~06382+1017, originally identified toward SFO~25 by \citet{sugitani1991catalog} using Palomar Sky
Survey red plates and the Infrared Astronomical Satellite (IRAS) Point Source Catalog, does not have a Gaia counterpart with reliable distance information and therefore cannot be used for this purpose. \citet{hosoya2020spectroscopic} identified four T~Tauri stars associated with our observed field of view of SFO~25 based on near-infrared color criteria derived from 2MASS photometry. Gaia distance estimates are available for all four sources satisfying the criteria RUWE $<1.4$; however, one object located at RA $=6^{\rm h}41^{\rm m}7.01^{\rm s}$ and Dec $=10^{\circ}16^{\prime}28.9^{\prime\prime}$ has a Gaia distance of $265 \pm 5$~pc, which is inconsistent with both the distance of HD~47839 and the molecular cloud complex. This source is therefore interpreted as a foreground object seen in projection and is excluded from our distance determination.

Several submillimetre studies have reported YSO candidates toward SFO~25. \citet{buckle2015wide} identified five 850~$\mu$m dust clumps using the Submillimetre Common-User Bolometer Array (SCUBA) on the JCMT telescope (sources 11, 12, 38, 73, and 130 in their Table~A1), of which three (38, 73, and 130) are classified as starless and are not considered further. The remaining two clumps (11 and 12), associated with Class~0/I protostars identified from Spitzer catalog of \cite{rapson2014spitzer}, do not have Gaia distance estimates. Similarly, the Class~I sources reported by \citet{wolf2003star} and the two Class~0 protostars identified by \citet{morgan2008scuba} using SCUBA observations, based on the criterion $L_{\rm submm}/L_{\rm bol} \gtrsim 5\times10^{-3}$ for Class 0 protostars \citep{andre1993submillimeter}, lack Gaia parallaxes. Consequently, YSOs identified exclusively from submillimetre observations are excluded from the distance analysis.

To augment the sample of YSOs with reliable distance information, we conducted an independent search using mid-infrared photometry from the \emph{AllWISE} \textbf{\citep{cutri2013vizier}} point source catalog. Sources were retrieved from the NASA/IPAC Infrared Science Archive (IRSA) within a radius of $7\arcmin$ centered on SFO~25 (RA $=6^{\rm h}41^{\rm m}6.91^{\rm s}$, Dec $=10^{\circ}17^{\prime}52.10^{\prime\prime}$). To ensure robust photometry, we required signal-to-noise ratios $\geq 5$ in the W1 (3.4~$\mu$m), W2 (4.6~$\mu$m), and W3 (12~$\mu$m) bands, which corresponds to the threshold corresponding to the source reliability criterion adopted in the AllWISE data processing and catalog construction. Sources affected by saturation, contamination or confusion (\texttt{cc\_flags} = `0000'), or extended emission (\texttt{ext\_flg} = 0) were excluded, restricting the sample to high-quality point sources. Of the 817 initially detected sources, 59 satisfied these quality criteria.

YSO candidates were identified using the color-based classification scheme of \citet{koenig2011wide}, based on their locations in the $W1-W2$ versus $W2-W3$ color--color diagram constructed from WISE profile-fitting magnitudes. Objects with $W1-W2 > 1.0$ and $W2-W3 > 2.0$ were classified as Class~I protostars, indicative of deeply embedded sources with substantial circumstellar envelopes. Class~II candidates, representing disk-bearing pre-main-sequence stars, were identified using the more conservative criteria $(W1-W2) - \sigma_{W1-W2} > 0.25$ and $(W2-W3) - \sigma_{W2-W3} > 1.0$, excluding sources already classified as Class~I. Applying these criteria, we identified one Class~I and five Class~II YSO candidates, while the remaining sources were classified as non-YSO objects.

Of the six newly identified YSOs, only two Class~II objects fall within the observed field of view of our SFO~25 molecular-line observations. Among them, only one source, located at RA $=6^{\rm h}41^{\rm m}9.74^{\rm s}$ and Dec $=10^{\circ}21^{\prime}36.15^{\prime\prime}$, has a reliable Gaia parallax measurement and satisfies the RUWE $<1.4$ criterion. The spatial distribution of all previously reported YSOs physically associated with SFO 25 and newly identified YSOs are shown in Figure~\ref{Fig: star formation}. Gaia parallaxes, corresponding inferred distances, proper motions in right ascension ($\mu_{\rm ra}$) and declination ($\mu_{\rm dec}$), and the radial velocities ($v_{\rm rad}$) are listed in Table~\ref{Table: table YSO} for those YSOs with available and reliable Gaia measurements.

Using the Gaia distances of YSOs deemed physically associated with SFO~25, we derive a mean distance to the cloud of $748.3 \pm 23.0$~pc. Their Gaia proper motions are also similar, with a dispersion of less than $\sim1$ mas yr$^{-1}$ in both right ascension and declination. At the distance of the region, this corresponds to a tangential velocity difference of only 2--3 km s$^{-1}$. In addition, the available radial velocities for these sources lie in the range 21--25 km s$^{-1}$, indicating a small velocity spread. The consistency in distance and kinematic properties suggests that these YSOs are likely associated with the SFO~25 cloud. Therefore, we adopt a representative distance of $\sim748$ pc for the cloud in the present analysis. At this distance, the line-of-sight separation between SFO~25 and HD~47839 is approximately 35~pc. This separation is consistent with the expected geometry of radiation-driven implosion systems and comparable to that observed in other RDI candidates, such as the L1616 cloud, where the dominant ionizing star HD~37128 lies at a distance of $60 \pm 81$~pc \citep{ramesh1995study, saha2022investigation, porel2025investigating}.

\begin{table*}
\centering
\caption{Positions, Gaia-based distances, proper motions, and radial velocities of the YSOs toward SFO~25.}
\label{Table: table YSO}
\renewcommand{\arraystretch}{1.3}
\begin{tabular}{lcccccccc}
\hline
YSO & Reference & RA (J2000) & Dec (J2000) & Parallax & Distance & $\mu_{ra}$ & $\mu_{dec}$ & v$_{rad
}$ \\
 &  & ($^{\rm h}\ ^{\rm m}\ ^{\rm s}$) 
 & ($^{\circ}\ ^{\prime}\ ^{\prime\prime}$) 
 & (mas) & (pc) & mas yr$^{-1}$ & mas/yr$^{-1}$ & km s$^{-1}$\\
\hline
LkHA~46 
& \cite{hosoya2020spectroscopic} 
& 06 41 01.53  
& +10 14 56.1 
& $1.40 \pm 0.07$ 
& $713.8 \pm 33.7$ & -1.79 & -3.48 & 24.95 \\

ESO-HA~493  
& \cite{hosoya2020spectroscopic} 
& 06 41 07.48 
& +10 15 04.5  
& $1.36 \pm 0.12$ 
& $734.5 \pm 65.3$ & -1.18 & -3.51 & ---- \\

ESO-HA~504  
& \cite{hosoya2020spectroscopic} 
& 06 41 09.85 
& +10 15 02.6 
& $1.25 \pm 0.18$ 
& $799.4 \pm 113.6$ & -2.01 & -4.01 & 23.27 \\

J064109.74+102136.1  
& This work 
& 06 41 09.74 
& +10 21 36.15 
& $1.29 \pm 0.06$ 
& $778.4 \pm 37.1$ & -1.44 & -3.33 & 21.01 \\
\hline
\end{tabular}
\end{table*}

\subsection{Cloud properties in the head and tail part of SFO 25}
\label{section: Cloud properties in the head and tail part of SFO 25}

\subsubsection{Excitation temperature and optical depth}
\label{section: tex and tau}

\begin{figure*}
\begin{center}
\resizebox{16.0cm}{11.0cm}{\includegraphics{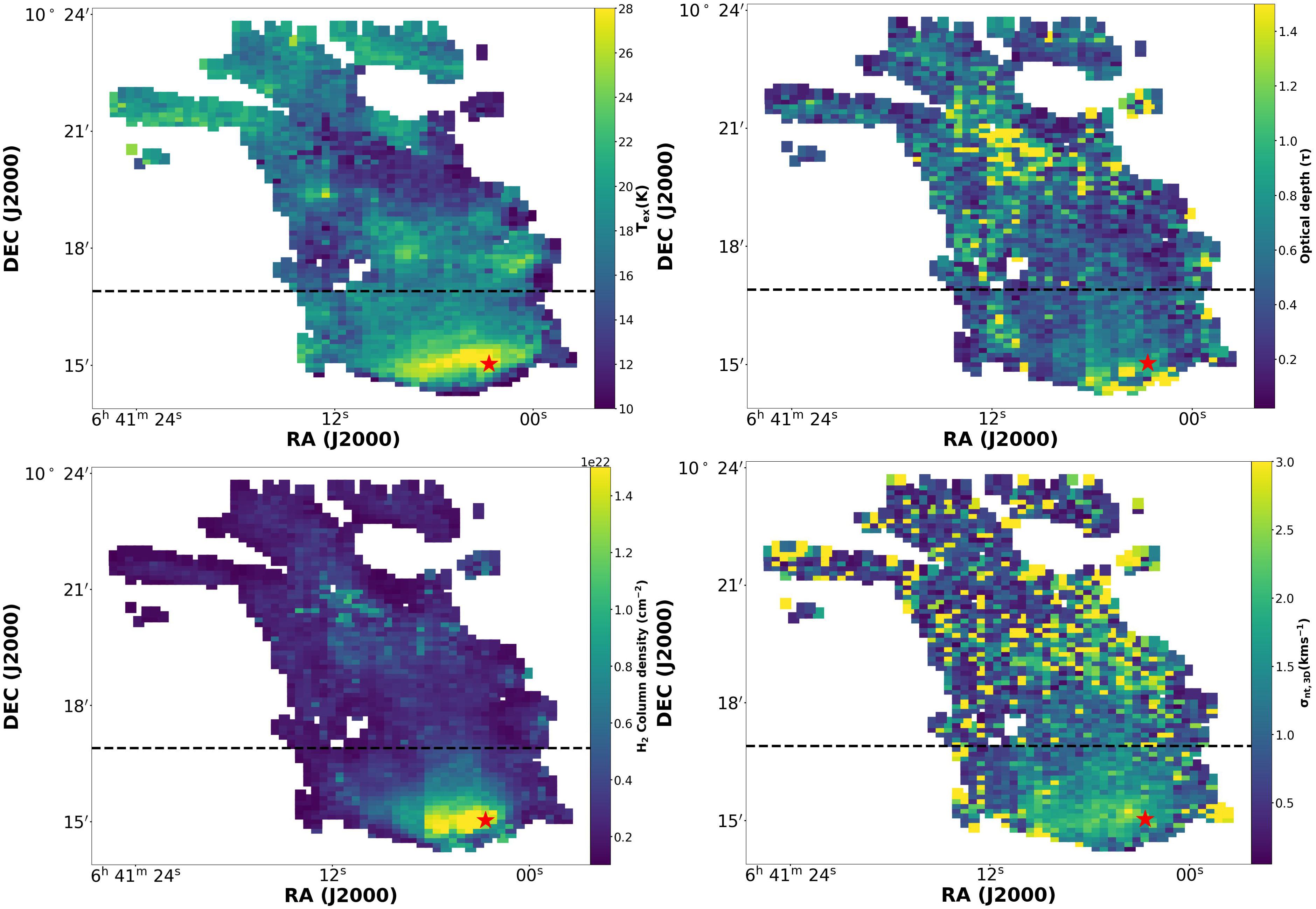}}
\caption{\textbf{Upper Left:} Excitation temperature ($T_{\rm ex}$) distribution. 
\textbf{Upper Right:} Optical depth ($\tau$) map. 
\textbf{Lower Left:} Molecular hydrogen column density ($N_{\rm H_2}$) map derived from $^{13}$CO (3--2) emission. 
\textbf{Lower Right:} Three-dimensional non-thermal velocity dispersion ($\sigma_{\rm nt,\,3D}$) map, also based on $^{13}$CO (3--2) emission. 
All panels represent the spatial distribution of key physical parameters within the region traced by $^{13}$CO (3--2) emission. 
The red stars indicate the position of the associated IRAS point source. The black dashed line delineates the boundary between the dense head and the elongated tail of the BRC SFO~25.
}\label{Fig: tex tau cd sigma map}
\end{center}
\end{figure*}

Assuming the condition of local thermodynamic equilibrium (LTE), the excitation temperature ($T_{ex}$) of the molecular gas was derived from the peak main beam temperature of the optically thick $^{12}$CO (3--2) line emission using the following relation \citep{pineda2008co, buckle2010jcmt}:
\begin{equation}
    T_{ex}^{12} \left(3 \rightarrow 2\right) = \frac{16.59\,\text{K}}{\ln\left[1 + \frac{16.59\,\text{K}}{T_{mb}^{\text{peak}} + 0.036\,\text{K}}\right]},
    \label{eq: 12CO tex equation}
\end{equation}
where $T_{mb}^{peak}$ represents the peak main-beam temperature, derived through Gaussian fitting of the $^{12}$CO spectral profile. This formulation is valid under the assumption that the $^{12}$CO emission is not significantly affected by self-absorption. However, in regions where $^{12}$CO exhibits prominent self-absorption, the $^{13}$CO line can become optically thick, often exhibiting a higher peak main beam temperature than $^{12}$CO. In such cases, the excitation temperature is estimated using the $^{13}$CO (3--2) transition via the following expression \citep{porel2025investigating, porel2025unbound}:
\begin{equation}
    T_{ex}^{13} \left(3 \rightarrow 2\right) = \frac{15.89\,\text{K}}{\ln\left[1 + \frac{15.89\,\text{K}}{T_{mb}^{\text{peak}} + 0.044\,\text{K}}\right]},
    \label{eq: 13CO tex equation}
\end{equation}
where $T_{mb}^{\text{peak}}$ is derived from the Gaussian fit to the $^{13}$CO spectral profile. Utilizing Equations~\ref{eq: 12CO tex equation} and \ref{eq: 13CO tex equation}, we constructed a pixel-by-pixel excitation temperature map over the entire $^{13}$CO emission region, as delineated by the outermost contour at approximately 1.68~K~km~s$^{-1}$.

The upper-left panel of Figure~\ref{Fig: tex tau cd sigma map} illustrates the spatial distribution of the derived excitation temperature. Within the head of the cloud, $T_{\rm ex}$ spans a range of 8.4--31.4~K, yielding a mean value of approximately 19.2~K. In comparison, the excitation temperatures within the tail region vary between 9.7--26.8~K with an average of about 16.8~K. Although both the head and tail are subjected to irradiation from the same ionizing source as discussed in detail in Section~\ref{section: Kinematics and Radiative Feedback in the SFO 25 Bright-Rimmed Cloud}, a notable distinction emerges in the head: a localized enhancement in excitation temperature is clearly concentrated around the IRAS source. This elevated $T_{\rm ex}$ is most plausibly attributed to internal energetic processes associated with the IRAS object---such as embedded heating or protostellar feedback---rather than external UV illumination alone, which otherwise influences the broader head and tail regions in a relatively uniform manner.
 
The optical depth of the $^{13}$CO (3--2) emission at line center, $\tau_0^{13}$, is estimated using the following expression \citep{porel2025unbound}:
\begin{equation}
    \tau_{0}^{13} = -\ln\left(1 - \frac{T^{13}_{mb,\text{peak}}}{15.89} \left[\frac{1}{\exp\left(\frac{15.89}{T_{ex}}\right) - 1} - \left(2.79 \times 10^{-3}\right) \right]^{-1} \right),
    \label{eq: 13co tau equation}
\end{equation}
where $T^{13}_{mb,\text{peak}}$ is the peak main beam temperature of the $^{13}$CO spectral profile and $T_{ex}$ is the excitation temperature previously derived for each pixel. Using Equation~\ref{eq: 13co tau equation}, we computed the optical depth on a pixel-by-pixel basis across the entire $^{13}$CO emission region as shown in the upper right panel of Figure~\ref{Fig: tex tau cd sigma map}.

In the head region of the cloud, the optical depth ranges from 0.03 to 2.94, with an average value of approximately 0.58. In the tail region, $\tau_0^{13}$ varies between 0.02 and 3.29, with a mean value of about 0.57. These results indicate that the $^{13}$CO emission remains predominantly optically thin in both regions. 

Table~\ref{Table: table 2} shows the range and mean values of excitation temperature for both head and tail regions of SFO 25 cloud.
Table~\ref{Table: table 1} presents the mean optical depth ($\bar\tau_{0}^{13}$), the full width at half maximum (FWHM; $\Delta V = 2.35\,\sigma_{\mathrm{obs}}$), and the three-dimensional velocity dispersion ($\sigma_{3D} = \sqrt{3}\,\sigma_{\mathrm{obs}}$) derived from Gaussian fits to the averaged spectral profiles of the head and tail regions in $^{13}$CO emission, as shown in Figure~\ref{Fig: head, tail of 13CO}. The parameter $\sigma_{\mathrm{obs}}$ corresponds to the one-dimensional velocity dispersion obtained from the Gaussian fitting procedure, with associated uncertainties estimated from the covariance matrix of the fit. The elevated values of FWHM and $\sigma_{3D}$ in the head region are attributed to enhanced turbulent motions, likely driven by outflow activity linked to class 0 or I type YSOs within this area.

\subsubsection{Mass and number density}
\label{section: mass and number density}

Assuming a circular morphology for the head region of SFO 25, we estimated its effective radius using the relation: $r_{\mathrm{eff}} = \sqrt{{A}/{\pi}}$
where $A$ denotes the projected area of the head region. The effective radius was determined to be approximately 0.36~pc.

In contrast, the tail region exhibits an elongated morphology, making the circular approximation inadequate. To account for this geometry, we employed the elliptical characterization method first introduced by \cite{stobie1980application} and subsequently applied by \cite{patel1995large}. Based on this approach, the derived geometrical parameters of the tail are a length of $L = 2a \approx 1.66$~pc and a width of $W = 2b \approx 1.10$~pc, where $a$ and $b$ denote the semi-major and semi-minor axes of the fitted ellipse, respectively.

Given the excitation temperature, we estimated the $^{13}$CO column density in both head and tail regions using the standard expression \citep{hayashi1991rp, buckle2010jcmt}:

\begin{equation}
    N\left(^{13}\mathrm{CO}\right) = 8.28 \times 10^{13} \exp\left(\frac{15.87}{T_{\mathrm{ex}}}\right) 
    \times \frac{T_{\mathrm{ex}} + 0.88}{1 - \exp\left(-\frac{15.87}{T_{\mathrm{ex}}}\right)} \int \tau\,dv.
    \label{eq: 13co column density}
\end{equation}
The integrated optical depth is related to the observed main-beam temperature through:

\begin{equation}
    \int \tau\,dv = \frac{1}{J\left(T_{\mathrm{ex}}\right) - J\left(T_{\mathrm{bg}}\right)} \int T_{\mathrm{mb}}\,dv \quad \text{for } \tau < 1,
    \label{eq: tau < 1}
\end{equation}

\begin{equation}
    \int \tau\,dv = 
    \frac{1}{J\!\left(T_{\mathrm{ex}}\right) - J\!\left(T_{\mathrm{bg}}\right)} 
    \cdot 
    \frac{\tau}{1 - e^{-\tau}} 
    \int T_{\mathrm{mb}}\,dv,
    \quad \text{for } \tau \geq 1.
    \label{eq:tau_greater_than_1}
\end{equation}
where the Planck function $J(T)$ is defined as:
\begin{equation}
    J(T) = \frac{h\nu/k}{\exp\left(h\nu/kT\right) - 1}.
\end{equation}

Once the $^{13}$CO column density is obtained, the corresponding H$_{2}$ column density can be estimated using the conversion \citep{frerking1982relationship, rawat2024giant}:
\begin{equation}
    N\left(\mathrm{H}_{2}\right) = 7 \times 10^{5} \times N\left(^{13}\mathrm{CO}\right).
    \label{eq: h2 column density relation}
\end{equation}

The lower left panel of Figure~\ref{Fig: tex tau cd sigma map} presents the spatial distribution of H$_{2}$ column density derived from $^{13}$CO emission across the entire $^{13}$CO-emitting region. A pronounced enhancement in column density is evident near the location of the IRAS source, marked by the red star symbol, suggesting the presence of a denser molecular concentration in this vicinity. Outside this localized high-density zone, both the head and tail regions exhibit relatively smooth variations in column density, with no significant pixel-to-pixel fluctuations, indicating a more uniform gas distribution across these parts.

The mass is determined using the following relation:
\begin{equation}  
    M = \mu_{\mathrm{H_{2}}}\, m_{\mathrm{H}}\, A_{\mathrm{pixel}} \sum N\left(\mathrm{H}_{2}\right),
    \label{eq: mass relation}
\end{equation}
where $\mu_{\mathrm{H_{2}}}$ is the mean molecular weight per hydrogen molecule, adopted as 2.8 following \citet{kauffmann2008mambo}, $m_{\mathrm{H}}$ is the mass of a hydrogen atom, and $A_{\mathrm{pixel}}$ denotes the physical area of a single pixel in units of cm$^{2}$. Using this equation, we estimated the masses of the head and tail regions to be 46~M$_\odot$ and 59~M$_\odot$, respectively.

The hydrogen number density ($n_{\mathrm{H_{2}}}$) is calculated using the following relations:
\begin{equation}
    n_{\mathrm{H_2}} = \frac{3M}{4\mu_{\mathrm{H_2}}\, m_{\mathrm{H}}\, \pi\, r_{\mathrm{eff}}^{3}},
    \label{eq: head and tail number density}
\end{equation}
where for the tail, the effective radius is $\sqrt{ab}$. Based on these calculations, the mean hydrogen number densities in the head and tail are estimated to be 3.5~$\times$~10$^{3}$~cm$^{-3}$ and 0.7~$\times$~10$^{3}$~cm$^{-3}$, respectively. This indicates that the tail region is significantly more diffuse compared to the head.

To evaluate the gravitational stability of the head and tail regions of SFO~25, we estimated
their virial masses using \citep{rawat2024giant, rawat2023probing}
\begin{equation}
M_{\mathrm{vir}} = 126 \left( \frac{5 - 2\beta}{3 - \beta} \right) r_{\mathrm{eff}} (\Delta V)^2,
\label{equation: virial mass}
\end{equation}
which accounts for non-uniform density
distributions. We adopt a density index of $\beta = 2$ for the approximately spherical head
and $\beta = 1.5$ for the elongated tail, consistent with power-law density profiles of the
form $\rho \propto r^{-\beta}$. The velocity dispersion, $\Delta V$, represents the mean line
width derived from pixel-by-pixel Gaussian fits to the $^{13}$CO data cube.

Using this formalism, we obtain virial masses of $199 \pm 11~M_\odot$ for the head and
$384 \pm 20~M_\odot$ for the tail. The corresponding virial parameters, defined as
$\alpha = M_{\mathrm{vir}}/M$, are $4.0 \pm 0.4$ and $6.0 \pm 0.5$, respectively. These values
indicate that both the head and the tail are gravitationally unbound.

To further assess the dynamical state of these regions and explicitly account for the influence of external radiation, we performed a complementary energy budget analysis incorporating gravitational, kinetic, and external pressure terms.
The external pressure was adopted from \citet{morgan2004radio}, who estimated the ionized
boundary layer (IBL) pressure toward SFO~25 using 20~cm (1.4~GHz) radio continuum emission
from the NRAO VLA Sky Survey (NVSS). Assuming optically thin free--free (thermal
bremsstrahlung) emission, they derived an ionized gas pressure of
$P_{\rm IBL}/k_{\rm B} = 6.64 \times 10^{6}$~cm$^{-3}$~K.

The gravitational potential energy of each region was computed following
\citet{zhou2023high}:
\begin{equation}
    W = -\frac{3}{5}\, a_{1} a_{2} \frac{G M^{2}}{r_{\rm eff}},
    \label{equation: potential energy}
\end{equation}
where $a_{1}$ accounts for the effect of a non-uniform density distribution and $a_{2}$
corrects for deviations from spherical symmetry. The factor $a_{1}$ is given by
$(1-\beta/3)/(1-2\beta/3)$, while $a_{2} = \arcsin(e)/e$, with the ellipticity defined as
$e = \sqrt{1-(b/a)^2}$. For the head region, a spherical geometry was assumed, with $e = 0$ and thus the $a_{2} \rightarrow 1$. In contrast, the tail exhibits a distinctly elongated morphology
and was therefore treated as an ellipsoidal structure, adopting an
ellipticity derived from its observed semi-major ($a$) and semi-minor ($b$) axes.

The kinetic energy was calculated as \citep{buckle2010jcmt}:
\begin{equation}
    E_{\rm kin} = \frac{1}{2} M \sigma_{3\rm D}^{2},
    \label{equation: kinetic energy}
\end{equation}
where $\sigma_{3\rm D}$ represents the three-dimensional velocity dispersion. To minimize
potential overestimation, $\sigma_{3\rm D}$ was computed as the spatial average over all
pixels within the respective head and tail regions. The resulting gravitational potential
and kinetic energies, summarized in Table~\ref{Table: table 2}, are of order
$10^{37}$~erg and $10^{38}$~erg, respectively, for both components.

The contribution of external pressure to the virial balance was estimated following \citet{zhou2023high}:
\begin{equation}
E_{\rm ext} = -4\pi P_{\rm ext} r_{\rm eff}^{3},
\label{eq: external energy}
\end{equation}
where $P_{\rm ext}$ corresponds to the pressure of the ionized boundary layer (IBL). Since a well-defined IBL is detected only in front of the head of SFO~25, the external pressure was estimated for this region alone. Although the tail appears to experience radiative acceleration (Section~\ref{section: Kinematics and Radiative Feedback in the SFO 25 Bright-Rimmed Cloud}), no distinct ionized layer is observed there, likely because the photoevaporated gas from the tail is highly diffuse and of low emission measure. Consequently, such diffuse ionized gas is unlikely to provide sufficient pressure to support or confine the molecular material in the tail. The derived external energy for the head is of order $10^{34}$~erg, more than three orders of magnitude smaller than the corresponding kinetic energy.

Including the external pressure term, the virial parameter was computed as
\begin{equation}
\alpha = \frac{2E_{\rm kin}}{|W| + |E_{\rm ext}|}.
\label{eq: virial parameter from energy}
\end{equation}
We obtain $\alpha \simeq 4.28$ for the head, indicating that this region remains gravitationally unbound even when the contribution of ionized gas pressure is taken into account. These results reinforce the conclusion that internal motions dominate the energy balance of SFO~25, and that neither the head nor the tail is expected to undergo large-scale gravitational collapse at present.

Table~\ref{Table: table 2} summarizes the minimum, maximum, and mean values of the column density, and effective radius, mass, and number density derived for the head and tail regions within SFO 25.

\subsubsection{Turbulent properties}
\label{section: turbulent properties}

To probe the turbulent characteristics within the head and tail regions of the SFO 25, we derived a suite of key dynamical parameters, including the thermal and non-thermal velocity dispersions and the Mach number. The thermal velocity dispersion (\(\sigma_{\mathrm{th}}\)) was calculated using the following expression:

\begin{equation}
    \sigma_{\mathrm{th}} = \sqrt{\frac{k_B T}{\mu_i m_H}},
    \label{eq: thermal velocity dispersion}
\end{equation}
where T is taken as the excitation temperature, and \(\mu_i = 29\) is the molecular weight of $^{13}$CO \citep{rawat2024giant}.

The one-dimensional non-thermal velocity dispersion (\(\sigma_{\mathrm{nt, 1D}}\)) was computed as:

\begin{equation}
    \sigma_{\mathrm{nt, 1D}} = \sqrt{\sigma_{\mathrm{obs}}^2 - \sigma_{\mathrm{th}}^2},
    \label{eq: non thermal 1d velocity dispersion}
\end{equation}
which quantifies the excess broadening in the observed line width beyond thermal contributions. The corresponding three-dimensional non-thermal velocity dispersion (\(\sigma_{\mathrm{nt, 3D}}\)) is given by:
$\sigma_{\mathrm{nt, 3D}} = \sqrt{3} \, \sigma_{\mathrm{nt, 1D}}$.

The thermal sound speed (\(c_s\)) was estimated using:

\begin{equation}
    c_s = \sqrt{\frac{k_B T}{\mu m_H}},
    \label{eq: thermal sound speed}
\end{equation}
where \(\mu = 2.37\) \citep{kauffmann2008mambo} represents the mean molecular weight per free particle in the molecular gas \citep{kauffmann2008mambo}. 

The degree of turbulence is characterized by the Mach number (\(\mathcal{M}\)), defined as:

\begin{equation}
    \mathcal{M} = \frac{\sigma_{\mathrm{nt, 3D}}}{c_s}.
    \label{eq: Mach number}
\end{equation}

The lower right panel of Figure~\ref{Fig: tex tau cd sigma map} presents the spatial distribution of the three-dimensional non-thermal velocity dispersion within the SFO~25 cloud, as traced by the $^{13}$CO (3--2) emission. Notably, the uncertainty in $\sigma_{\mathrm{nt,\,3D}}$ is relatively higher toward the tail region, reflecting the increased noise level in this part of the cloud. Table~\ref{Table: table 1} lists the mean values of the thermal velocity dispersion, 3D non-thermal velocity dispersion, sound speed, and Mach number for the head and tail subregions. The mean Mach numbers for both components are comparable within 2$\sigma$, indicating that the turbulent motions are strongly supersonic throughout the cloud.

\subsection{Inside the Clump: Probing the Physical Conditions of a Star-Forming Region}
\label{section: Inside the Clump: Probing the Physical Conditions of a Star-Forming Region}

\begin{figure*}
\begin{center}
\resizebox{16.0cm}{6.5cm}{\includegraphics{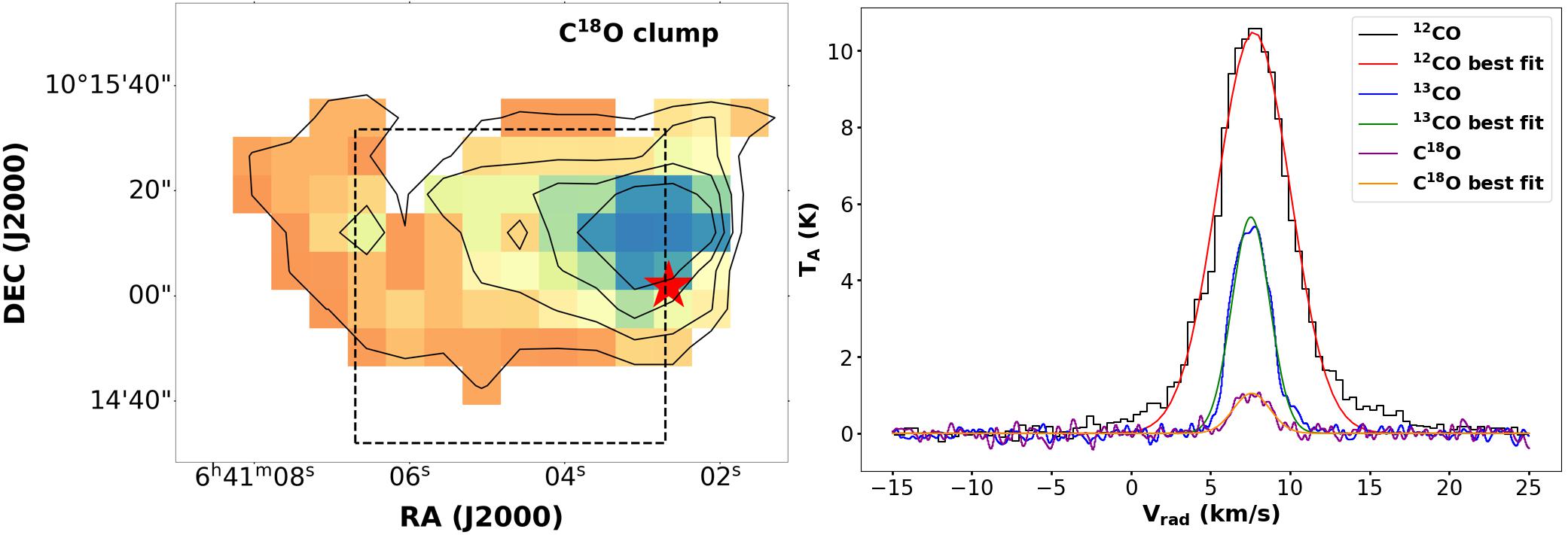}}
\caption{\textbf{Left:} Integrated intensity (moment 0) map of C$^{18}$O (3--2) emission constructed over the velocity range of 5.02 to 10.08 km s$^{-1}$. The map is overlaid with contours representing C$^{18}$O integrated intensities exceeding the 3$\sigma$ noise level, where $\sigma \approx 0.69$~K~km~s$^{-1}$ corresponds to the rms noise in the background. Contour levels are drawn in steps of 1$\sigma$. The red star marks the position of the associated IRAS source. The dashed black box outlines the approximate extent of the core region mapped in N$_2$H$^{+}$ emission, as presented in \citet{de2002star}. 
\textbf{Right:} Molecular line spectra extracted from the clump region in $^{12}$CO (3--2) (black), $^{13}$CO (3--2) (blue), and C$^{18}$O (3--2) (purple) transitions. The corresponding Gaussian fits to the spectra are overplotted in red, green, and orange, respectively, highlighting the velocity components and line profiles of each tracer.
}
\label{Fig: sfo 25 clump}
\end{center}
\end{figure*}

As C$^{18}$O is a well-established tracer of dense molecular gas, it serves as an effective probe for identifying compact structures such as clumps within molecular clouds. Utilizing this tracer, we identified a prominent clump associated with the SFO 25 bright-rimmed cloud, located in the southern head region of the structure. This clump hosts the embedded IRAS source. The left panel of Figure~\ref{Fig: sfo 25 clump} displays the identified clump in the C$^{18}$O (3--2) moment 0 map, integrated over the velocity range of 5.02 to 10.08 km~s$^{-1}$. The map is overlaid with contours representing integrated intensities above the 3$\sigma$ level, where $\sigma$ denotes the standard deviation of the background noise.

The right panel of Figure~\ref{Fig: sfo 25 clump} presents the spectral profiles extracted from the clump region in $^{12}$CO, $^{13}$CO, and C$^{18}$O (3--2) emission, shown in black, blue, and purple, respectively. Overlaid Gaussian fits are plotted in red, green, and orange, respectively. The systemic velocities derived from these tracers are 7.72 km~s$^{-1}$ for $^{12}$CO, 7.54 km~s$^{-1}$ for $^{13}$CO, and 7.59 km~s$^{-1}$ for C$^{18}$O, all of which indicate that the clump is blue-shifted relative to the dominant red-shifted velocity component of the entire $^{13}$CO-emitting cloud (8.87 km s$^{-1}$). 

Table~\ref{Table: table 1} summarizes the mean values of $\bar\tau_{0}^{18}$, line width, $\sigma_{3D}$, and V$_{\mathrm{LSR}}$ derived from C$^{18}$O emission. For comparison, if we calculate $\Delta V$ and $\sigma_{3D}$ within this clump using $^{13}$CO emission, the values obtained are 2.79~km~s$^{-1}$ and 2.06~km~s$^{-1}$, respectively. These values are significantly higher than those observed in the head and tail regions. This enhancement may be attributed to the presence of an embedded IRAS source, known to exhibit strong molecular outflow activity \citep{sugitani1991catalog, de2002star}, which likely injects energy into the surrounding gas, thereby increasing the velocity dispersion and the associated dynamical parameters within the clump.

The excitation temperature within the clump varies between 19.1--31.4~K, with the mean value around 25.5~K. The optical depth of the C$^{18}$O emission ranges from 0.02 to 0.46, with a mean value of approximately 0.16. This optical depth was calculated using the following relation \citep{porel2025investigating, porel2025unbound}:

\begin{equation}
    \tau_{0}^{18} = -\ln\left(1 - \frac{T^{18}_{\mathrm{mb},\text{peak}}}{15.79} \left[\frac{1}{\exp\left(\frac{15.79}{T_{\mathrm{ex}}}\right) - 1} - \left(2.89 \times 10^{-3}\right) \right]^{-1} \right).
    \label{eq: c18o tau equation}
\end{equation}

Given that C$^{18}$O traces dense material effectively, the hydrogen column density was derived from the C$^{18}$O column density using the following relation \citep{hayashi1991rp, buckle2010jcmt}:

\begin{equation}
    N\left(\mathrm{C}^{18}\mathrm{O}\right) = 8.26 \times 10^{13} \exp\left(\frac{15.81}{T_{\mathrm{ex}}}\right) 
    \times \frac{T_{\mathrm{ex}} + 0.88}{1 - \exp\left(-\frac{15.81}{T_{\mathrm{ex}}}\right)} \int \tau\,dv,
    \label{eq: c18o column density}
\end{equation}
and subsequently \citep{frerking1982relationship},
\begin{equation}
    (N\left(\mathrm{H}_2\right))_{\mathrm{C}^{18}\mathrm{O}} = 7 \times 10^{6} \, N\left(\mathrm{C}^{18}\mathrm{O}\right).
    \label{eq: h2 cd based on c18o}
\end{equation}
Application of these expressions yields a hydrogen column density ranging from 1.1~$\times$~10$^{22}$ to 2.5~$\times$~10$^{22}$~cm$^{-2}$, with a mean value of approximately 1.6~$\times$~10$^{22}$~cm$^{-2}$. Assuming a circular geometry and an effective radius of $\sim$0.13~pc (calculated similarly to the head effective radius), we estimate the mass and number density of the clump to be 18~M$_\odot$ and 2.9~$\times$~10$^{4}$~cm$^{-3}$, respectively. This indicates that the clump is significantly denser—by nearly an order of magnitude—compared to the average conditions in the broader head region.

In the study by \citet{de2002star}, a dense core was identified in N$_2$H$^+$ emission—a well-known tracer of high-density gas ($\gtrsim 10^5\ \mathrm{cm}^{-3}$)—as illustrated in their Figure 19. In our work, the approximate extent of this mapped core region is represented by the dashed black box in the left panel of Figure~\ref{Fig: sfo 25 clump}. Although \citet{de2002star} did not provide derived structural parameters for this core, we estimate its effective radius to be $\sim0.12\ \mathrm{pc}$, which closely matches the effective radius we derive for the C$^{18}$O clump.

Within this mapped core, \citet{de2002star} measured an excitation temperature of approximately 5 K from the N$_2$H$^+$ line—significantly lower than our average excitation temperature for the clump obtained from $^{12}$CO emission. This temperature disparity likely reflects the differing density sensitivities of the tracers: $^{12}$CO preferentially traces the more diffuse, outer layers of the molecular gas, which are warmed by radiation from the ionizing star, while N$_2$H$^+$ traces the shielded, denser interior that remains colder. It should be noted, however, that \citet{de2002star} recognized that their excitation temperature estimate is poorly constrained due to noise.

Moreover, \citet{de2002star} estimated the core mass to be $\sim50\ M_{\odot}$, which is considerably higher than our derived clump mass. This discrepancy is consistent with the distinct gas density regimes probed by the two tracers—N$_2$H$^+$ is sensitive to densities $\gtrsim10^5\ \mathrm{cm}^{-3}$ \citep{de2002star}, whereas C$^{18}$O traces gas at densities $\gtrsim10^4\ \mathrm{cm}^{-3}$ \citep{buckle2010jcmt}—resulting in a greater inferred mass in the high-density regions traced by N$_2$H$^+$.

Regarding the gravitational stability of the identified C$^{18}$O clump, we derive a
virial mass of $M_{\rm vir} = 107 \pm 24~M_\odot$ and a corresponding virial parameter of
$\alpha \simeq 6 \pm 1$. Such an elevated virial parameter unambiguously indicates that the
clump is gravitationally unbound. This dynamically unbound state is plausibly enhanced by
feedback from the embedded IRAS source, whose associated outflow activity can inject
substantial kinetic energy into the surrounding gas, thereby increasing internal turbulence and promoting dispersal.

To further assess the robustness of this conclusion, we performed an independent energy
budget analysis incorporating gravitational potential energy, internal kinetic energy,
and the contribution of external pressure from the ionized boundary layer. Using
Equations~(\ref{equation: potential energy}), (\ref{equation: kinetic energy}), and
(\ref{eq: external energy}), and adopting an external pressure equal to that acting
on the head region, we find a virial parameter of $\alpha \simeq 5.8$. The persistence of
$\alpha > 2$, even after accounting for the confining effect by the expansion of the ionized gas,
demonstrates that external pressure is insufficient to stabilize the clump against
internal motions. We therefore conclude that the clump is not gravitationally bound and is
unlikely to undergo collapse under the current radiative and dynamical conditions.
 
Table~\ref{Table: table 2} summarizes the minimum, maximum, and mean values of the excitation temperature, column density, effective radius, mass, number density, gravitational
potential energy, and kinetic energy of the identified clump.

Table~\ref{Table: table 1} summarizes the mean values of the turbulent parameters within the clump region, derived from C$^{18}$O emission. The results indicate that the clump exhibits supersonic turbulent motion, which is consistent with the presence of active outflow from the embedded IRAS source. For the calculation of the thermal velocity dispersion, a molecular weight of $\mu_i = 30$ was adopted \citep{rawat2024giant}, corresponding to the C$^{18}$O tracer used in the analysis.

\begin{table*}
\begin{center}
    \caption{Physical properties and gravitational stability of the head and tail regions traced by $^{13}$CO emission, and of the clump region traced by C$^{18}$O emission within the SFO 25 cloud.}
    \label{Physical properties for SFO 38}
    \renewcommand{\arraystretch}{1.3} 
    \begin{tabular}{lccc ccc cc cc c} 
        \hline
        Region & \multicolumn{3}{c}{$T_{\rm kin}$ (K)} & \multicolumn{3}{c}{$N(H_2)$ ($\times10^{22}$ cm$^{-2}$)} & $r_{eff}$ & Mass & $\langle n(H_2)\rangle$ & -W & E$_{kin}$ \\
               & Min & Max & Mean & Min & Max & Mean & (pc) & ($M_\odot$) & ($\times10^3$ cm$^{-3}$) & ($\times 10^{37}$J) & ($\times 10^{37}$J) \\
        \hline
        Head & 8.4 & 31.4 & $19.2\pm0.2$ & 0.11 & 1.94 & $0.53\pm0.02$ & $0.36 \pm 0.01$ & $46\pm3$ & $3.5 \pm 0.4$ & $5.2 \pm 0.7$ & $11.2 \pm 0.8$ \\
        Tail & 9.7 & 26.8 & $16.8\pm0.1$ & 0.10 & 1.05 & $0.27\pm0.01$ & $0.68 \pm 0.01$ & $59\pm4$ & $0.7 \pm 0.1$ & $3.8 \pm 0.5$ & $11.0 \pm 0.9$ \\
        Clump & 19.1 & 31.4 & $25.5\pm0.3$ & 1.12 & 2.49 & $1.55\pm0.05$ & $0.13 \pm 0.004$ &  $18\pm1$ & $29.1 \pm 3.2$ & $2.2 \pm 0.3$ & $6.5 \pm 1.5$\\
        \hline 
    \end{tabular}
    \label{Table: table 2}
\end{center}
\end{table*}

\begin{table*}
\begin{center}
	\caption{Kinematic and turbulent properties of the head and tail regions determined from $^{13}$CO (3--2) emission, and of the dense clump region derived from C$^{18}$O (3--2) emission within the SFO 25. The values of $\Delta V$, $\sigma_{3D}$, and $V_{\mathrm{LSR}}$ are extracted from the averaged spectral profiles of the respective regions, as illustrated in Figures~\ref{Fig: head, tail of 13CO} and~\ref{Fig: sfo 25 clump}. The turbulent parameters represent mean values calculated over all spatial pixels encompassed within each region.}
    \renewcommand{\arraystretch}{1.3} 
	\begin{tabular}{lccccccccc} 
		\hline
		Region & $\bar \tau_{0}$ & $\Delta V$  & $\sigma_{3D}$ & $V_{LSR}$ & $\langle \sigma_{th} \rangle$ & $\langle \sigma_{nt, 3D} \rangle$ & $\langle c_{s} \rangle$ & $\langle \mathcal{M} \rangle$ \\ 
		  &  & (km s$^{-1}$) & (km s$^{-1}$) & (km s$^{-1}$) & (km s$^{-1}$) & (km s$^{-1}$) & (km s$^{-1}$) & \\ 
		\hline
    Head & $0.58 \pm 0.01$ & $2.28\pm0.02$ & $1.68\pm0.01$ & $7.65\pm 0.01$ & 0.07 & $1.54\pm0.03$ & 0.26 & $6.02\pm0.14$\\
    Tail & $0.57 \pm 0.01$ &  $1.69\pm0.03$ & $1.25\pm0.02$ & $8.79\pm0.01$ & 0.07 & $1.35\pm0.03$ & 0.24 & $5.69\pm0.15$\\
    Clump & $0.16 \pm 0.01$ &  $2.71\pm0.08$ & $2.00\pm0.06$ & $7.59\pm0.04$ & 0.08 & $1.48\pm0.09$ & 0.30 & $4.97\pm0.30$\\
		\hline 
	\end{tabular}
    \label{Table: table 1}
\end{center}
\end{table*}

\section{Discussion}
\label{section: discussion}

\begin{figure*}
\begin{center}
\resizebox{16.0cm}{6.5cm}{\includegraphics{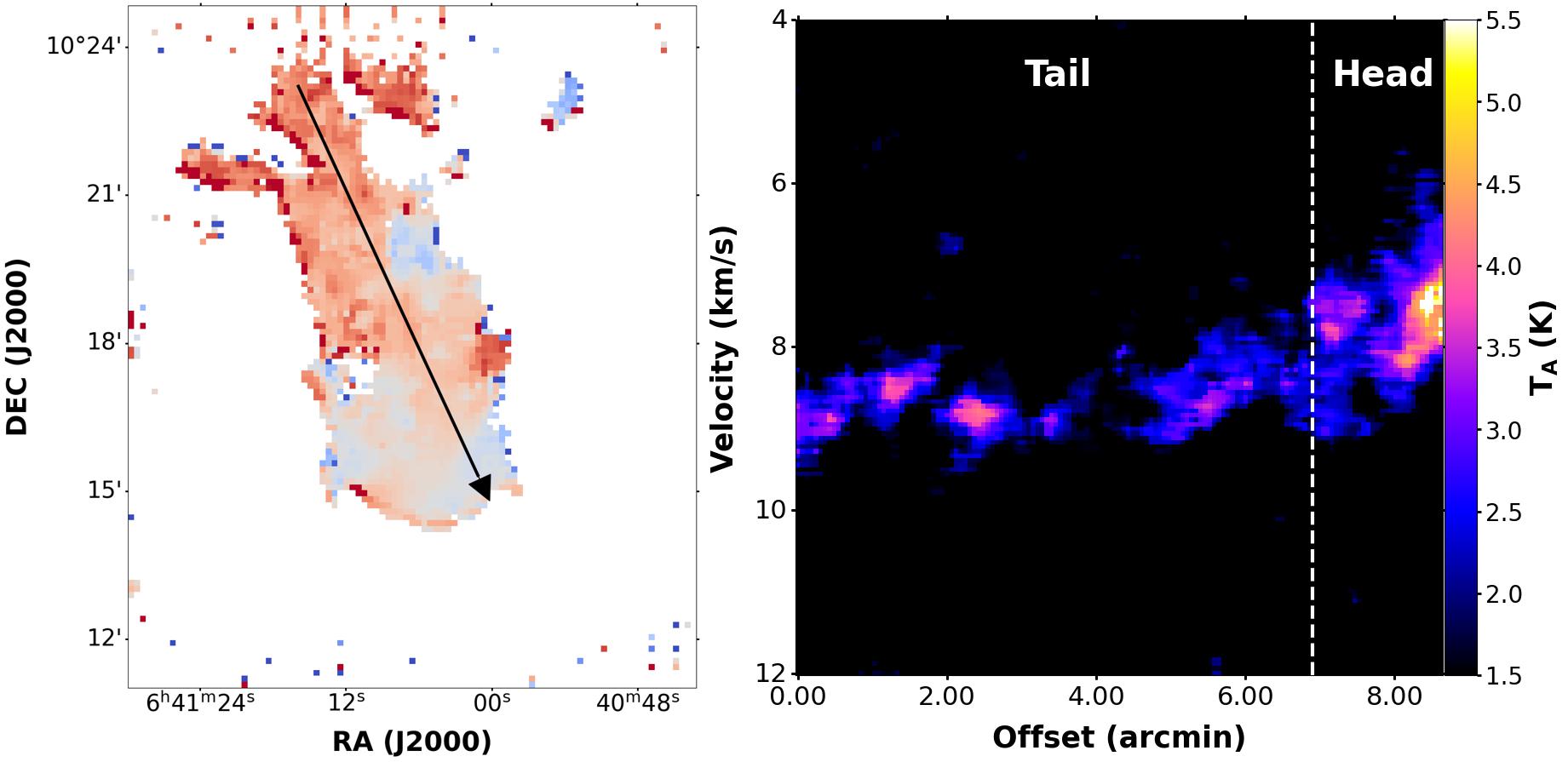}}
\caption{Position–velocity (PV) diagram obtained along a linear slice of length $8.68^{\prime}$ overlaid on the moment~1 map of the $^{13}$CO~($J=3$--2) emission, tracing the kinematic structure of the molecular gas along the selected direction.}
\label{Fig: sfo 25 pv}
\end{center}
\end{figure*}

The primary objective of this study is to assess the future star formation potential of the head and tail regions of the SFO~25 cloud, determining whether these structures can evolve into active sites of star formation or be dispersed under the influence of intense ionizing radiation. This goal is addressed through the quantitative evaluation of key physical parameters, including the virial analysis, and energy budget estimation derived in Section~\ref{section: mass and number density}, photoevaporative mass-loss rate, depletion timescale, and photoevaporative timescale. A detailed discussion of these aspects is presented in the following sections.

\subsection{Kinematics and Radiative Feedback in the SFO 25 Bright-Rimmed Cloud}
\label{section: Kinematics and Radiative Feedback in the SFO 25 Bright-Rimmed Cloud}

We investigate the kinematics and radiative influence on the SFO~25 bright-rimmed cloud in order to assess how external ultraviolet irradiation governs its structural
evolution and star-forming potential. Our analysis combines molecular-line kinematics,
photoevaporation-driven mass-loss estimates, and dynamical diagnostics. Together, these diagnostics provide a
coherent picture of the interaction between stellar feedback and the evolving cloud
structure.

The moment~1 map of $^{13}$CO~(3--2) emission (Figure~\ref{Fig: moment 1 map}) reveals a pronounced velocity gradient across SFO~25,
which is also evident in the corresponding position--velocity (PV) diagram
(Figure~\ref{Fig: sfo 25 pv}), oriented from the redshifted tail toward the blueshifted
head. \citet{buckle2012structure} derived a systemic velocity of $\sim$5.0~km~s$^{-1}$ for the adjacent H\,\textsc{ii} region NGC~2264G based on JCMT--HARP $^{12}$CO~(3--2) observations. In contrast, Gaussian fits to the $^{13}$CO spectra extracted from the head and tail regions of SFO~25 (Figure~\ref{Fig: head, tail of 13CO}) yield centroid velocities of $V_{\rm LSR} \approx 7.7$~km~s$^{-1}$ and $8.8$~km~s$^{-1}$, respectively. These results demonstrate that the molecular gas in SFO~25 is systematically redshifted with respect to the surrounding ionized environment. Consistent with this finding, the $^{12}$CO~(3--2) emission map (Figure~\ref{Fig: Regions in 12co}) shows that Region~1, which traces the main body of the SFO~25 cloud, is dominated by a redshifted velocity component centered near 8.8~km~s$^{-1}$. The close agreement between the velocity fields derived from both $^{13}$CO and $^{12}$CO indicates that the observed kinematic offset represents that the molecular gas of the SFO 25 cloud is redshifted with respect to the ionized gas in NGC 2264 H\,\textsc{ii} region and the tail is redshifted with respect to the head of SFO 25 bright-rimmed cloud.

Based on the study of \cite{zhou2025kinematics}, young stellar objects are expected to inherit the kinematic properties of their natal molecular clouds during the early stages of their evolution, although older pre-main-sequence stars may gradually decouple from the surrounding gas. In the case of SFO~25, Gaia data are available only for the T~Tauri stars located in the head and a single Class~II protostar situated in the tail, while no Gaia measurements are available for Class~0 or I protostars in the head region. To assess whether projection effects significantly influence the observed gas motions within the head and tail, we rely on the kinematics of these T~Tauri and Class~II YSOs.

From Gaia, we have obtained the right ascension and declination proper motions, parallaxes, and radial velocities for two T~Tauri stars, Lk-H$\alpha$~46 and ESO-H$\alpha$~504 in the head, and for one Class~II protostar in the tail (Table~\ref{Table: table YSO}). The 3D velocity of each star is calculated as $v_\mathrm{3D} = \sqrt{v_\mathrm{tan}^2 + v_\mathrm{rad}^2}$, where $v_\mathrm{tan}$ is the tangential velocity derived from proper motions. We define the angle $\theta$ between the 3D velocity vector and the line-of-sight (radial) velocity, related as $v_\mathrm{rad} = v_\mathrm{3D} \cos\theta$. In this framework, $\theta = 0^\circ$ corresponds to motion entirely along the line of sight, while $\theta = 90^\circ$ indicates motion entirely in the plane of the sky. Consequently, small values of $\theta$ ($< 45^\circ$) indicate that the radial velocity closely approximates the true 3D motion, implying a minimal to moderate projection effect.

In the tail region, the barycentric radial velocity of the gas ($\sim 24~\mathrm{km~s^{-1}}$) and the 3D velocity of the Class~II protostar from Gaia ($\sim 25~\mathrm{km~s^{-1}}$) yield $\theta \sim 16^\circ$, indicating that approximately 96\% of the gas motion is along the line of sight. In the head region, the radial gas velocity ($\sim 23~\mathrm{km~s^{-1}}$) compared with the 3D velocity of the T~Tauri stars ($\sim 28~\mathrm{km~s^{-1}}$ for Lk-H$\alpha$ 46 and $\sim$29 km s$^{-1}$ for ESO-H$\alpha$ 504) gives average $\theta \sim 36^\circ$, corresponding to roughly 81\% of the motion along the line of sight. The difference in $\theta$ between the head and tail likely reflects variations in the orientation of the velocity vectors in 3D space, as well as the limited sample of stars, rather than indicating significant projection effects. These results indicate that the radial velocities in both regions closely capture the true 3D motions of the gas, suggesting that the observed gas motions might be genuine and not strongly affected by projection effects. We note, however, that these conclusions are based on a limited sample—one YSO in the tail and two YSOs in the head and the YSOs are also T Tauri and class II type—and should therefore be considered indicative rather than definitive. Nevertheless, the derived $\theta$ values and corresponding line-of-sight fractions provide supporting evidence that the radial velocities can serve as reliable tracers of the underlying gas kinematics. Accordingly, the observed redshift of the diffuse tail relative to the head might reflect actual gas motion, which is further supported by the PV-diagram, where a clear velocity gradient is seen from tail to head.

We note, however, that a rigorous confirmation of minimal projection effects in the head region would require Gaia measurements for the Class~0 or I protostars, which are presently unavailable. For the tail, such data are not obtainable, as Class~0 or I protostars are not present in this region.

Based on Gaia DR3 astrometry \citep{vallenari2023gaia}, the ionizing O7V star HD~47839
is located at a distance of $\sim 713$~pc, while SFO~25 lies farther along the line of sight at $\sim 748$~pc. This configuration, together with the systemic velocity estimation of the molecular gas in SFO 25 cloud with respect to the ionized gas in NGC 2264 H\,\textsc{ii} region and the minimal projection effect might imply that the cloud is situated behind the ionizing source, such that ultraviolet radiation from HD~47839 impinges upon both the head and tail regions of SFO~25. The observed redshifted motion of the tail with respect to the head at present may therefore reflect the dynamical response of the more diffuse tail material to the incident radiation. Owing to its comparatively lower density and weaker gravitational confinement, the tail is likely more susceptible to photoevaporative effects. Consequently, the tail material can experience a larger acceleration under the influence of radiation-driven forces, as estimated later in this section. It is worth noting that this behavior does not necessarily imply that the radiation field is stronger at the tail than at the head; rather, the enhanced response is likely a consequence of the lower density and more diffuse structure of the tail, which allows the incident radiation to more efficiently accelerate and disperse the material.

The observed tail-to-head velocity gradient does not imply mass transport from the diffuse tail toward the comparatively dense head. Instead, the measured systemic velocities indicate that the tail is receding with respect to the head along the line of sight. This behavior is consistent with a scenario in which external radiative feedback drives bulk motion and gradual erosion of the cloud material.

To quantify the role of photoevaporation, we adopt the formulation of
\citet{codella2001star}:

\begin{equation}
    \dot{M}_{\rm loss} =
    29 \left(\frac{\Phi_{LyC}}{10^{7}\, \rm cm^{-2}\, s^{-1}}\right)^{1/2}
    \left(\frac{r_{eff}}{1\, \rm pc}\right)^{3/2}
    \, M_\odot\, \rm Myr^{-1},
\end{equation}
where $\Phi_{LyC} = N_{L}/4\pi d^{2}$ is the Lyman–continuum photon flux incident upon the cloud from HD 47839, $N_{L}$ is the Lyman–continuum photon rate for O7V type star adopted as 10$^{48.86}$ s$^{-1}$ \citep{panagia1973some} and $d$ is the projected distance from the ionizing star to the head or tail regions. Applying this prescription yields photoevaporative mass--loss
rates of $\dot{M}_{\rm head} \sim 112~M_\odot~\rm Myr^{-1}$ and
$\dot{M}_{\rm tail} \sim 229~M_\odot~\rm Myr^{-1}$, consistent with the presence of
extended ionized gas traced by H$\alpha$ emission
(Figure~\ref{Fig: wise and optical image}, middle panel). Using the masses derived in
Section~\ref{section: mass and number density}, the corresponding instantaneous gas depletion timescales, defined as $t_{\rm dep} = M/\dot{M}_{\rm loss}$, are $\sim 0.42$~Myr for the head and $\sim 0.26$~Myr for the tail under the assumption of uniform mass loss rate. These results indicate that photoevaporative mass removal is presently more efficient in the tail, which is expected given its lower density and greater susceptibility to external ionizing radiation due to its redshifted gas motion, implying that the tail will be dispersed on a shorter timescale than the head.

We estimate the approximate radiative force acting on the head and tail of SFO~25 as a consequence of photoevaporation using the relation $F_{\rm rad} \sim \dot{M}_{\rm loss} c_{s,i}$, where $c_{s,i}$ denotes the sound speed of the ionized gas (Equation~\ref{eq: thermal sound speed}) and $\dot{M}_{\rm loss}$ is the photoevaporative mass--loss rate. For the ionized gas, we assume a characteristic temperature of $T = 10^{4}$ K \citep{morgan2004radio} and adopt a mean molecular weight $\mu = 0.61$ \citep{spitzer2008physical}. Using these parameters, we obtain radiative forces of $F_{\rm tail} \approx 2.69 \times 10^{3}$ $M_\odot$ km s$^{-1}$ Myr$^{-1}$ for the tail and $F_{\rm head} \approx 1.31 \times 10^{3}$ $M_\odot$ km s$^{-1}$ Myr$^{-1}$ for the head. The tail therefore experiences approximately twice the radiative force compared to the head, which is consistent with its comparatively diffuse structure and lower density, making it more susceptible to the dynamical influence of radiation-driven flows.

A similar behavior has been reported for the southern head and the north–eastern (NE) tail of the bright-rimmed cloud SFO~38 \citep{porel2025unbound, choudhury2010triggered, patel1995large}. In that region, the NE tail is observed to be redshifted relative to the head (see Table~3 of \citealt{porel2025unbound}), and previous studies suggest that this tail portion is strongly influenced by ionizing radiation from the O-type star HD~206267 \citep{porel2025unbound, choudhury2010triggered}. Using the same formulation, we estimate photoevaporative mass–loss rates of $\sim 82~M_\odot \ {\rm Myr^{-1}}$ for the head and $\sim 101~M_\odot \ {\rm Myr^{-1}}$ for the NE tail. The corresponding radiative forces are $\sim 0.96 \times 10^{3}$ $M_\odot \ {\rm km \ s^{-1}\ Myr^{-1}}$ and $\sim 1.19 \times 10^{3}$ $M_\odot \ {\rm km \ s^{-1} \ Myr^{-1}}$ for the head and NE tail, respectively. These values indicate that the more diffuse NE tail experiences a stronger radiative force than the head, which may contribute to its larger redshifted motion and enhanced mass-loss rate.

In both bright-rimmed clouds, the ratio of the incident Lyman–continuum photon flux between the head and tail regions is relatively modest—approximately 1.4 for SFO~25 and about 1.2 for SFO~38. This indicates that although the head receives marginally higher ionizing photon flux than the tail, the difference is not substantial. Consequently, the comparatively diffuse nature of the tail allows the incident radiation to accelerate and remove material more efficiently than in the denser head region. This behavior provides a consistent explanation for the enhanced radiative force, larger mass-loss rate, and the observed redshifted motion of the tail in both BRC systems.



Virial parameter and energy budget analyses
(Sections~\ref{section: mass and number density} and
\ref{section: Inside the Clump: Probing the Physical Conditions of a Star-Forming Region})
demonstrate that both the head and the identified C$^{18}$O clump are gravitationally
unbound. As a consequence, the dynamical evolution of SFO~25 is dominated by external
radiative erosion feedback rather than self-gravity, rendering large-scale gravitational collapse
unlikely.

Although the head is comparatively denser than the tail, its gravitationally unbound state and rapid photoevaporative erosion likewise argue against future star formation. Previous studies by \citet{morgan2004radio} showed that the compact N$_2$H$^+$ core identified by \citet{de2002star} is already subject to strong external compression, with ionized-gas pressure exceeding internal turbulent support. Because C$^{18}$O traces gas at lower critical densities than N$_2$H$^+$, the structure identified in our analysis represents a more extended, moderately dense component rather than a bound prestellar core.

To assess the prospects for future star formation, we adopt the instantaneous gas depletion
timescale, which directly compares the available
molecular gas reservoir with the current rate of photoevaporative mass removal. The inferred
depletion timescales for both the head and tail regions are significantly shorter than the
characteristic timescales for core formation or gravitational collapse, even under the
conservative assumption that these structures are gravitationally bound and evolve in free
fall $t_{\rm ff} = \sqrt{\frac{3\pi}{32 G \rho}}$ timescale, where $\rho$ is the mass density. This implies that radiative feedback will disperse the
molecular gas before gravity can act efficiently to trigger star formation. 

\subsection{Star Formation Status and Potential in SFO 25}
\label{section: Star Formation Status and Potential in SFO 25}

Figure~\ref{Fig: star formation} shows the spatial distribution of YSOs overlaid on the $^{13}$CO moment~0 map, with C$^{18}$O integrated intensity contours shown in black. Several YSOs exhibit extremely small angular separations. In particular, YSO~25~NIR identified by \citet{wolf2003star} lies $\sim$0.89$\arcsec$ from the IRAS source; YSO~25~S1 \citep{wolf2003star} and JCMTSF~J064104.2+093507 \citep{buckle2015wide} are separated by $\sim$1.07$\arcsec$; and YSO~25~S2 \citep{wolf2003star} and JCMTSF~J064103.5+101508 \citep{buckle2015wide} are separated by $\sim$2.73$\arcsec$. These separations are significantly smaller than the pixel size of our data (7.3$\arcsec$), which naturally accounts for their apparent spatial coincidence in the map.

The spatial distribution of YSOs demonstrates that star formation activity in SFO~25 is strongly concentrated toward the dense head of the bright-rimmed cloud. As discussed in Section~\ref{section: Distance to SFO 25 and Its Physical Association with HD 47839}, the sources identified by \citet{wolf2003star} and the associated IRAS object are classified as Class~I YSOs, those reported by \citet{morgan2008scuba} correspond to Class~0 protostars, and the sources identified by \citet{buckle2015wide} belong to the Class~0/I category. In addition, the stars identified by \citet{hosoya2020spectroscopic} are T~Tauri stars. The close association of these deeply embedded YSOs with the densest molecular gas in the head region provides strong evidence for recent star formation history in this part of the cloud, consistent with previous studies of SFO~25.

In contrast, earlier investigations did not report evidence of star formation in the tail region. Using the ALLWISE \citep{cutri2013vizier} catalog, we identify two Class~II YSOs located within the tail (Section~\ref{section: Distance to SFO 25 and Its Physical Association with HD 47839}). This new detection indicates that star formation in SFO~25 might not be restricted solely to the dense head but has also occurred within the lower-density tail. Furthermore, the distinct evolutionary classes of YSOs found in the two regions, the presence of Class~0, I protostars in the head and only more evolved Class~II sources in the tail—likely reflect the underlying density contrast between these components. The higher column density and volume density of the head appear to sustain more recent star formation activity, whereas the more diffuse tail is characterized by an older and less vigorous star-forming episode.

\cite{sugitani1991catalog} proposed that radiation-driven implosion may have triggered star formation in the head of the bright-rimmed cloud SFO~25. This interpretation was based on the ratio of infrared luminosity ($L_{\rm IR}$, derived from the IRAS 12, 25, 60, and 100~$\mu$m bands) of the IRAS source located immediately behind the bright rim to the mass of the cloud head. This ratio was found to be significantly larger than the typical value of $\sim 0.1~L_\odot\,M_\odot^{-1}$ observed in isolated Bok globules. Such an enhanced $L_{\rm IR}/M_{\rm cloud}$ ratio is widely regarded as an observational signature of externally triggered star formation induced by ionizing radiation.

The ionizing star HD~47839, the dominant source of the cluster NGC 2264 \citep{dahm2008young}, is assumed to have an age approximately $1.63^{+3.27}_{-0.27}$~Myr \citep{sung2010initial}. The older stellar populations within SFO~25 include T Tauri stars in the head and Class II protostars in the tail \citep{hosoya2020spectroscopic}. To assess whether RDI could plausibly have triggered star formation in these regions, it is essential to estimate the ages of these comparatively older young stellar objects. Due to the lack of cataloged ages in Vizier, we derived approximate ages using Gaia DR3 photometry and astrometry for the T Tauri stars in the head and the Class II protostar in the tail for which Gaia data are available.

The Gaia G-band magnitudes and parallaxes of stars in the head and tail regions of SFO~25, used for distance determination (Section~\ref{section: Distance to SFO 25 and Its Physical Association with HD 47839}), enable independent approximate age estimation. Absolute magnitudes were calculated using $M_G = G + 5 \log_{10}(\varpi) - 10 - A_G$, where $\varpi$ is the parallax in milliarcseconds and $A_G$ is the extinction in the Gaia G band. Extinction was estimated for each star based on the column density at its position in the cloud, converted to $A_V$, and then to $A_G$ using standard extinction relations. Intrinsic colors $(BP-RP)_0$ were obtained after correcting for reddening. The extinction-corrected color–magnitude diagram (CMD; $M_G$ versus $(BP-RP)_0$) was then compared with pre–main-sequence (PMS) isochrones from \cite{baraffe2015new} to infer stellar ages via interpolation between the nearest bracketing isochrones. This method provides approximate ages with uncertainties of a few Myr, primarily arising from parallax, photometric, and extinction corrections, as well as model assumptions.

Using this approach, we find that the Class II star in the tail region has an approximate age of $\sim 5 \pm 1$ Myr. In contrast, T Tauri stars in the head region exhibit significantly younger ages: Lk-H$\alpha$ ($\sim 0.5$–$1$ Myr), ESO-H$\alpha$ 493 ($\sim 0.5$–$1$ Myr), and ESO-H$\alpha$ 504 ($\sim 1$–$2$ Myr). While these age estimates are approximate and assume solar metallicity, they are broadly consistent with expectations for T Tauri stars or class II protostars.

According to the radiation-driven implosion models of \cite{bisbas2011radiation}, the minimum time required for the formation of the first protostar after the onset of external ionizing radiation is given by $t_* = 0.19~{\rm Myr}\,(\Phi_{\rm LyC}/10^{9}\,{\rm cm^{-2}\,s^{-1}})^{-1/3}$. Here, $t_*$ denotes the delay between the activation of the ionizing source and the onset of gravitational collapse within the irradiated cloud, and therefore represents the minimum expected age difference between the ionizing star and the first star formed as a consequence of RDI.

The estimated age difference between the ionizing star HD~47839 ($\sim 1.4$--4.9~Myr) and the oldest T Tauri star in the head region ($\sim 1$--2~Myr) is therefore of order $\sim 0.4$--3~Myr. Since from the radiation-driven implosion models of \cite{bisbas2011radiation}, a minimum triggering timescale of $t_* \sim 0.13$~Myr for SFO 25 is needed, this observed age gap is compatible with the expected delay between the onset of ionizing radiation and the formation of the first protostar. Although the stellar ages are larger than the theoretical triggering timescale, these quantities describe different evolutionary stages: $t_*$ corresponds to the onset of protostellar collapse, whereas the observed ages reflect subsequent pre--main--sequence evolution. The observed age sequence, in which the low-mass stars are younger than the ionizing source, is therefore qualitatively consistent with the expectations of RDI within the substantial uncertainties associated with PMS age determinations, extinction corrections, and the adopted evolutionary models. Consequently, the age relationship provides supportive evidence for RDI as a plausible mechanism for triggering star formation in the head of SFO~25 in the earlier evolutionary stage of this cloud.

In contrast, the star located in the tail region appears to be older than HD~47839, which argues against RDI as the triggering mechanism for star formation in the tail. However, since our stellar ages are derived from Gaia-based color--magnitude diagram analysis rather than from homogeneous cataloged values, these conclusions remain subject to uncertainties arising from photometry, extinction corrections, and evolutionary model assumptions.

A similar situation was reported by \cite{makela2012star} for the cometary globule CG~1, where star formation was identified in both the head and tail regions. While RDI was proposed as the likely triggering mechanism in the head, the origin of star formation in the tail could not be conclusively explained.

The classical radiation-driven implosion framework describes the interaction between ionizing radiation and a molecular cloud under an idealized geometry in which the dense head directly faces the ionizing source, while the tail remains shielded. In this scenario, ionization of the cloud surface produces a photoevaporative flow that drives a compressive shock into the head, potentially triggering star formation. The expected kinematic signature is that the tail appears blueshifted relative to the head along the cloud axis due to the expansion of ionized gas away from the irradiated surface \citep{bisbas2011radiation, sugitani1991catalog, makela2012star}. Observational evidence supporting such behavior has been reported in several cometary clouds (e.g., \citealt{saha2022investigation, makela2012star}).

The velocity structures traced in the $^{12}$CO and $^{13}$CO moment~1 maps of SFO~25, along with the position--velocity diagram, the discernible extension of H$\alpha$ emission into portions of the tail beyond the dense head (middle panel of Figure~\ref{Fig: wise and optical image}) and the derived radiative force values (considering minimal projection effects), suggest that both the head and the tail are exposed to ionizing radiation from HD~47839. This indicates that the tail may not be effectively shielded by the head, contrary to the assumptions of an idealized radiation-driven implosion geometry.

In the SFO~38 cloud, the kinematic behavior of different substructures is non-uniform. The NE tail and the southern head exhibit a velocity distribution similar to that observed in SFO~25, where the tail is more redshifted than the head because the NE tail of SFO~38 is exposed to ionizing radiation from HD~206267 as suggested by \cite{choudhury2010triggered, patel1995large, porel2025unbound, okada2024bright}, indicating a deviation from the classical RDI picture. In contrast, the north-western (NW) tail is more blueshifted than the head, which is consistent with the classical RDI scenario \citep{porel2025unbound, patel1995large}.


In SFO~25, projection effects are found to be small, implying that the deviation from the classical RDI kinematic pattern is intrinsic. This suggests that factors such as complex cloud morphology, asymmetric density distribution, and non-uniform irradiation may play a dominant role in shaping the velocity field. A similar interpretation may also apply to SFO~38, where geometric and environmental complexities are likely to contribute to the observed non-uniform kinematic behavior \citep{okada2024bright}.

Despite these deviations, the star formation activity in the dense heads of both SFO~38 and SFO~25 is consistent with an RDI-triggered origin, where compression by photoevaporative flows induces gravitational collapse. The current physical states of the two clouds, however, indicate different evolutionary stages. The head of SFO~38 remains gravitationally bound, suggesting that it may continue forming stars, whereas the head of SFO~25 appears to be gravitationally unbound, indicating that further star formation may not be sustained.

Additionally, the NE tail of SFO~38 shows no evidence of star formation, while the tail of SFO~25 does exhibit star-forming activity. However, this activity is unlikely to be driven by the RDI mechanism, and its origin remains uncertain. Moreover, given that the tail of SFO~25 appears to be gravitationally unbound, it is unlikely to sustain further star formation in the future.

Overall, these results indicate that deviations from the classical RDI velocity structure might not necessarily rule out RDI-triggered star formation in dense cloud heads, but instead reflect intrinsic structural complexity and evolutionary effects.

In this work, all derived physical parameters—including excitation temperature, optical depth, column density, mass, number density, as well as the virial parameter and the associated gravitational and kinetic energies—are estimated under the assumption of local thermodynamic equilibrium. This approximation is widely employed in recent studies of star-forming and feedback-dominated molecular cloud environments, including Orion-B \citep{buckle2010jcmt}, where JCMT-HARP observations of the CO $J=3\rightarrow2$ transition ($^{12}$CO, $^{13}$CO, and C$^{18}$O) were analysed, as well as the G148.24+00.41 hub–filament system \citep{rawat2024giant}, investigated using the Milky Way Imaging Scroll
Painting (MWISP) survey data from the Purple Mountain Observatory (PMO) 13.7-m telescope in the CO isotopologues $J=1\rightarrow0$ transitions. Similar LTE-based analyses have also been performed for photodissociation region and bright-rimmed cloud, such as L1616 and SFO 38, respectively \citep{porel2025investigating, porel2025unbound}, using JCMT-HARP observations of the CO isotopologues $J=3\rightarrow2$ lines, despite these regions being influenced by radiative feedback, outflows, and dynamical compression.

Furthermore, dedicated comparisons between LTE and non-LTE approaches indicate that LTE can provide a reasonable first-order description under typical molecular cloud conditions. In particular, \cite{zhou2024high} performed a combined LTE and RADEX-based non-LTE analysis of the G333 giant molecular cloud (GMC) using high-resolution Atacama Pathfinder Experiment/Large APEX sub-Millimeter Array (APEX/LAsMA) observations of the CO $J=3\rightarrow2$ transition (primarily $^{12}$CO and $^{13}$CO), and found that for hydrogen number densities of order $\sim 4.2 \times 10^{3}\,\mathrm{cm^{-3}}$, the LTE-derived quantities are broadly consistent with the non-LTE solutions. Motivated by these results, and by the consistency reported in such CO (3–2) studies, we adopt the LTE approximation as a first-order approach for the physical characterization of SFO 25 bright-rimmed cloud.

However, we acknowledge that the LTE assumption may not be strictly valid across the entire region of SFO 25. Recent investigations by \cite{barnes2020lego} have demonstrated that molecular line emission cannot be uniquely associated with gas at or above the critical density, and that the commonly assumed one-to-one correspondence between critical density and the traced volume density is an oversimplification. In the sub-millimetre regime, deviations from LTE arise naturally when the gas density is below the effective excitation threshold, leading to sub-thermal excitation and excitation temperatures lower than the kinetic temperature. In the framework of recent numerical and theoretical studies \citep{barnes2020lego}, molecular line emission is not a direct tracer of gas mass at a single characteristic density, but is instead weighted toward regions of parameter space where excitation, optical depth, and abundance jointly maximise the line emissivity, implying that LTE-based mass estimates reflect an emission-weighted subset of the gas rather than the total mass distribution. In such regimes, LTE-based column densities and masses may be systematically biased low. Conversely, local enhancements in temperature, radiation field, or shock-driven compression can increase excitation, introducing additional departures from LTE conditions. More generally, line emission is governed by a combination of density, temperature, optical depth, molecular abundance, and radiative transfer effects along the line of sight, rather than a single controlling parameter such as critical density. For a more comprehensive discussion of the limitations of the LTE approximation and the interpretation of molecular line emission in contemporary theoretical and observational frameworks, we refer the reader to \citet{barnes2020lego}.

For the CO $J=3\rightarrow2$ transitions analysed here, the characteristic critical density is $\sim10^{4}-10^{5}\,\mathrm{cm^{-3}}$ for typical cloud temperatures of 10–50 K \citep{buckle2010jcmt}. The hydrogen number densities inferred for both the head and tail regions are generally below this range, suggesting that sub-thermal excitation may be present. Under these conditions, the LTE-derived excitation temperatures might be interpreted as upper limits, while the corresponding column densities and masses as well as the gravitational energy, kinetic energy, and the Mach numbers might represent lower limits. This propagates directly into the virial analysis: although the virial mass is independent of LTE assumptions, the virial parameter is sensitive to the LTE-derived mass and is therefore likely overestimated. In this context, achieving virial equilibrium would require masses of approximately 100 $M_{\odot}$ for the head and 256 $M_{\odot}$ for the tail, corresponding to factors of $\sim2$ and $\sim4$ higher than the LTE-derived values, respectively. Hence, any significant non-LTE correction increasing the true mass would reduce the inferred virial parameter and could alter the dynamical classification toward a more gravitationally bound state.

A fully self-consistent non-LTE analysis would ideally combine multiple CO transitions, including previously observed lower-$J$ data (e.g., $J=1\rightarrow0$ and $J=2\rightarrow1$ from FCRAO and HHT \citep{de2002star}) with the JCMT $J=3\rightarrow2$ observations used in this work. Such a multi-transition approach, implemented through radiative transfer modelling (e.g., RADEX), would allow tighter constraints on the kinetic temperature, density, and optical depth structure. However, this is presently limited by differences in angular resolution, spatial coverage, beam filling factors, and calibration consistency among available datasets. A robust non-LTE treatment would therefore require new, spatially matched multi-line observations, which is beyond the scope of the present study.

In addition, the present analysis adopts a single characteristic gas density derived in Sections~\ref{section: mass and number density} and~\ref{section: Inside the Clump: Probing the Physical Conditions of a Star-Forming Region}. We acknowledge that this assumption represents a simplification of the actual physical conditions, as molecular-line emission in molecular clouds is generally expected to arise from gas spanning a range of densities rather than from a single homogeneous density component. Future observations covering a larger number of molecular transitions will not only enable more robust non-LTE analyses but also allow the application of radiative-transfer methods that explicitly account for multi-density gas distributions (e.g., the Dense Gas Toolbox; \citealt{puschnig2020dense}). Such approaches have the potential to provide more realistic constraints on the density distribution and excitation conditions of the molecular gas, leading to a more comprehensive interpretation of the physical conditions within the cloud.

Independent constraints from dust continuum measurements are also limited. While SCUBA observations have identified two cold cores in the head of SFO 25 with dust temperatures of around $19$~K \citep{morgan2008scuba}, broadly consistent with our inferred gas temperature in head, this does not establish thermal coupling across the full extent of the $^{13}$CO-emitting region. Furthermore, no sufficiently resolved Herschel-based dust temperature map is available for this source, preventing a direct comparison between gas and dust temperatures on cloud scales.

Despite the limitations inherent in the LTE approximation, the principal conclusions of this study are not expected to change substantially. In particular, the inferred kinematic characteristics of SFO~25, which indicate that the cloud may deviate from the classical RDI scenario, are based primarily on the observed relative velocity structure between the head and tail, the spatial distribution of the H$\alpha$ emission, and estimates of the radiative force. These diagnostics are largely independent of the LTE-derived physical parameters and are therefore relatively insensitive to plausible non-LTE effects. Furthermore, although the star formation activity in the head may have been triggered by RDI, this interpretation is independently supported by the relative ages of the young stellar objects and the ionizing star and is thus not expected to be significantly affected by uncertainties associated with the LTE assumption. Consequently, the overall conclusion that SFO~25 may exhibit departures from the predictions of the classical RDI scenario, as discussed in Sections~\ref{section: Kinematics and Radiative Feedback in the SFO 25 Bright-Rimmed Cloud} and \ref{section: Star Formation Status and Potential in SFO 25}, remains qualitatively unchanged under reasonable non-LTE corrections.

Future multi-transition, high-resolution observations combined with non-LTE radiative transfer modelling will be essential to refine the absolute physical parameters and further test the dynamical state of SFO 25. Nonetheless, within the uncertainties inherent to the LTE framework, our analysis provides a consistent first-order physical and kinematic characterization of the system.

\section{Conclusions}
\label{section: conclusion}

We investigated the morphology, kinematics, and dynamical state of the bright-rimmed cloud SFO~25 using archival JCMT--HARP observations of $^{12}$CO, $^{13}$CO, and C$^{18}$O ($J=3\rightarrow2$). We re-estimated the distance to SFO~25 using Gaia parallaxes of young stellar objects physically associated with the cloud, which place it behind the ionizing O7V star HD~47839 along the line of sight and found that the molecular gas of SFO 25 cloud is redshifted with respect to the ionized gas in the NGC 2264 H\,\textsc{ii} region. These results together with the consideration of minimal projection effect, might imply that ultraviolet radiation directly impinges upon both the head and tail of the cloud. The molecular gas exhibits a pronounced head--tail velocity gradient, with the tail systematically redshifted relative to the head, indicating that the present-day kinematics of SFO~25 deviate from the predictions of the classical radiation-driven implosion scenario. 

Owing to its more diffuse structure, the tail experiences stronger radiative acceleration than the head, a result that is supported by the derived radiative force values and photoevaporative mass-loss rates for the two regions. We identify signatures of star formation in the tail for the first time, a feature not reported in previous studies. While earlier works suggested that star formation in the head was influenced by radiation-driven implosion at the early stage of evolution of the SFO 25 BRC, our analysis provides additional observational support for this interpretation. In contrast, the physical mechanism responsible for the star formation activity in the tail remains uncertain, as indicated by the Gaia-based age estimates of the associated young stellar objects.  

Despite the presence of embedded Class~0/I protostars in the head, its current gravitationally unbound state implies that it is unlikely to sustain future star formation. The tail, which is both dynamically receding from the head and gravitationally unbound, is likewise expected to undergo continued dispersal, inhibiting further star formation within this region. Together, these results portray SFO~25 as an evolved bright-rimmed cloud whose present evolution is dominated by radiative dispersal rather than by gravitational collapse.

Our results indicate that, despite the deviation of the head--tail velocity structure from the classical RDI scenario, the star formation in the head is consistent with having been triggered by radiation-driven implosion.

\section*{Acknowledgements}

We thank the anonymous referee for their constructive comments and insightful suggestions, which have significantly improved the quality and clarity of this manuscript. We gratefully acknowledge the JCMT HARP data used in this study, obtained under project code M08BU15. The East Asian Observatory operates the JCMT in partnership with the National Astronomical Observatory of China, the National Astronomical Observatory of Japan, the Korea Astronomy and Space Science Institute (KASI), the Academia Sinica Institute of Astronomy and Astrophysics of Taiwan, and the Science and Technology Facilities Council of the United Kingdom.

We also make use of publicly available WISE 12 and 4 $\mu$m emission data, accessed through NASA’s \textit{SkyView Virtual Observatory} (\url{https://skyview.gsfc.nasa.gov}), and AllWISE point source catalog which greatly supported the analysis presented in this work.

This research was supported by the Indian Institute of Astrophysics (IIA) under the Department of Science and Technology (DST), Government of India.

\section*{Data Availability}

The observational data used in this study were obtained with the James Clerk Maxwell Telescope (JCMT), employing the Heterodyne Array Receiver Programme (HARP) to observe the $^{13}$CO and C$^{18}$O ($J=3\rightarrow2$) transitions. These data were retrieved from the JCMT Science Archive. Complementary infrared data from WISE, optical data, and H$\alpha$ emission from SHASSA were accessed through the NASA SkyView\footnote{\url{https://skyview.gsfc.nasa.gov/current/cgi/query.pl}} interface. In addition to the above datasets, we utilized data from the ALLWISE catalog.

The data analysis was performed using the \texttt{Python} ecosystem, utilizing key libraries including \texttt{Astropy} for handling astronomical data \citep{astropy2018astropy}, \texttt{SciPy} for numerical analysis \citep{virtanen2020scipy}, and \texttt{NumPy} for array-based computations \citep{harris2020array}.



\bibliographystyle{mnras}
\bibliography{example} 

\begin{thebibliography}{}
\makeatletter
\relax
\def\mn@urlcharsother{\let\do\@makeother \do\$\do\&\do\#\do\^\do\_\do\%\do\~}
\def\mn@doi{\begingroup\mn@urlcharsother \@ifnextchar [ {\mn@doi@} {\mn@doi@[]}}
\def\mn@doi@[#1]#2{\def\@tempa{#1}\ifx\@tempa\@empty \href {http://dx.doi.org/#2} {doi:#2}\else \href {http://dx.doi.org/#2} {#1}\fi \endgroup}
\def\mn@eprint#1#2{\mn@eprint@#1:#2::\@nil}
\def\mn@eprint@arXiv#1{\href {http://arxiv.org/abs/#1} {{\tt arXiv:#1}}}
\def\mn@eprint@dblp#1{\href {http://dblp.uni-trier.de/rec/bibtex/#1.xml} {dblp:#1}}
\def\mn@eprint@#1:#2:#3:#4\@nil{\def\@tempa {#1}\def\@tempb {#2}\def\@tempc {#3}\ifx \@tempc \@empty \let \@tempc \@tempb \let \@tempb \@tempa \fi \ifx \@tempb \@empty \def\@tempb {arXiv}\fi \@ifundefined {mn@eprint@\@tempb}{\@tempb:\@tempc}{\expandafter \expandafter \csname mn@eprint@\@tempb\endcsname \expandafter{\@tempc}}}

\bibitem[\protect\citeauthoryear{Andre, Ward-Thompson  \& Barsony}{Andre et~al.}{1993}]{andre1993submillimeter}
Andre P.,  Ward-Thompson D.,   Barsony M.,  1993, Astrophysical Journal, Part 1 (ISSN 0004-637X), vol. 406, no. 1, p. 122-141., 406, 122

\bibitem[\protect\citeauthoryear{Baraffe, Homeier, Allard  \& Chabrier}{Baraffe et~al.}{2015}]{baraffe2015new}
Baraffe I.,  Homeier D.,  Allard F.,   Chabrier G.,  2015, Astronomy \& Astrophysics, 577, A42

\bibitem[\protect\citeauthoryear{Barnes et~al.,}{Barnes et~al.}{2020}]{barnes2020lego}
Barnes A.~T.,  et~al., 2020, Monthly Notices of the Royal Astronomical Society, 497, 1972

\bibitem[\protect\citeauthoryear{Bertoldi}{Bertoldi}{1989}]{bertoldi1989photoevaporation}
Bertoldi F.,  1989, Astrophysical Journal, Part 1 (ISSN 0004-637X), vol. 346, Nov. 15, 1989, p. 735-755. Research sponsored by NASA., 346, 735

\bibitem[\protect\citeauthoryear{Bisbas, W{\"u}nsch, Whitworth, Hubber  \& Walch}{Bisbas et~al.}{2011}]{bisbas2011radiation}
Bisbas T.~G.,  W{\"u}nsch R.,  Whitworth A.~P.,  Hubber D.~A.,   Walch S.,  2011, The Astrophysical Journal, 736, 142

\bibitem[\protect\citeauthoryear{Buckle \& Richer}{Buckle \& Richer}{2015}]{buckle2015wide}
Buckle J.,  Richer J.,  2015, Monthly Notices of the Royal Astronomical Society, 453, 2006

\bibitem[\protect\citeauthoryear{Buckle et~al.,}{Buckle et~al.}{2010}]{buckle2010jcmt}
Buckle J.~a.,  et~al., 2010, Monthly Notices of the Royal Astronomical Society, 401, 204

\bibitem[\protect\citeauthoryear{Buckle, Richer  \& Davis}{Buckle et~al.}{2012}]{buckle2012structure}
Buckle J.,  Richer J.,   Davis C.,  2012, Monthly Notices of the Royal Astronomical Society, 423, 1127

\bibitem[\protect\citeauthoryear{Choudhury, Mookerjea  \& Bhatt}{Choudhury et~al.}{2010}]{choudhury2010triggered}
Choudhury R.,  Mookerjea B.,   Bhatt H.,  2010, The Astrophysical Journal, 717, 1067

\bibitem[\protect\citeauthoryear{Codella, Bachiller, Nisini, Saraceno  \& Testi}{Codella et~al.}{2001}]{codella2001star}
Codella C.,  Bachiller R.,  Nisini B.,  Saraceno P.,   Testi L.,  2001, Astronomy \& Astrophysics, 376, 271

\bibitem[\protect\citeauthoryear{Collaboration et~al.,}{Collaboration et~al.}{2018}]{astropy2018astropy}
Collaboration A.,  et~al., 2018, The Astronomical Journal, 156, 123

\bibitem[\protect\citeauthoryear{Cutri, Wright, Conrow  et~al.}{Cutri et~al.}{2013}]{cutri2013vizier}
Cutri R.,  Wright E.,  Conrow T.,   et~al., 2013, Originally published in: IPAC/Caltech

\bibitem[\protect\citeauthoryear{Dahm}{Dahm}{2008}]{dahm2008young}
Dahm S.,  2008, arXiv preprint arXiv:0808.3835

\bibitem[\protect\citeauthoryear{De~Vries, Narayanan  \& Snell}{De~Vries et~al.}{2002}]{de2002star}
De~Vries C.~H.,  Narayanan G.,   Snell R.~L.,  2002, The Astrophysical Journal, 577, 798

\bibitem[\protect\citeauthoryear{Elmegreen}{Elmegreen}{2011}]{elmegreen2011triggered}
Elmegreen B.~G.,  2011, EAS Publications Series, 51, 45

\bibitem[\protect\citeauthoryear{Frerking, Langer  \& Wilson}{Frerking et~al.}{1982}]{frerking1982relationship}
Frerking M.~A.,  Langer W.~D.,   Wilson R.~W.,  1982, Astrophysical Journal, Part 1, vol. 262, Nov. 15, 1982, p. 590-605. NASA-supported research., 262, 590

\bibitem[\protect\citeauthoryear{Harris et~al.,}{Harris et~al.}{2020}]{harris2020array}
Harris C.~R.,  et~al., 2020, Nature, 585, 357

\bibitem[\protect\citeauthoryear{Hayashi, Gatley, Hasegawa  \& Kaifu}{Hayashi et~al.}{1991}]{hayashi1991rp}
Hayashi M.,  Gatley I.,  Hasegawa T.,   Kaifu N.,  1991, THE ASTROPHYSICAL JOURNAL, 374, 540

\bibitem[\protect\citeauthoryear{Hosoya, Itoh, Oasa, Gupta  \& Sen}{Hosoya et~al.}{2020}]{hosoya2020spectroscopic}
Hosoya K.,  Itoh Y.,  Oasa Y.,  Gupta R.,   Sen A.~K.,  2020, arXiv preprint arXiv:2006.00219

\bibitem[\protect\citeauthoryear{Jenness \& Economou}{Jenness \& Economou}{2015}]{jenness2015orac}
Jenness T.,  Economou F.,  2015, Astronomy and Computing, 9, 40

\bibitem[\protect\citeauthoryear{Kauffmann, Bertoldi, Bourke, Evans  \& Lee}{Kauffmann et~al.}{2008}]{kauffmann2008mambo}
Kauffmann J.,  Bertoldi F.,  Bourke T.~L.,  Evans N.~J.,   Lee C.~W.,  2008, Astronomy \& Astrophysics, 487, 993

\bibitem[\protect\citeauthoryear{Koenig, Leisawitz, Benford, Rebull, Padgett  \& Assef}{Koenig et~al.}{2011}]{koenig2011wide}
Koenig X.,  Leisawitz D.,  Benford D.,  Rebull L.,  Padgett D.,   Assef R.,  2011, The Astrophysical Journal, 744, 130

\bibitem[\protect\citeauthoryear{Lee \& Chen}{Lee \& Chen}{2007}]{lee2007triggered}
Lee H.-T.,  Chen W.,  2007, The Astrophysical Journal, 657, 884

\bibitem[\protect\citeauthoryear{M{\"a}kel{\"a} \& Haikala}{M{\"a}kel{\"a} \& Haikala}{2012}]{makela2012star}
M{\"a}kel{\"a} M.,  Haikala L.,  2012, arXiv preprint arXiv:1211.1263

\bibitem[\protect\citeauthoryear{Morgan, Thompson, Urquhart, White  \& Miao}{Morgan et~al.}{2004}]{morgan2004radio}
Morgan L.~K.,  Thompson M.,  Urquhart J.,  White G.~J.,   Miao J.,  2004, Astronomy \& Astrophysics, 426, 535

\bibitem[\protect\citeauthoryear{Morgan, Thompson, Urquhart  \& White}{Morgan et~al.}{2008}]{morgan2008scuba}
Morgan L.,  Thompson M.,  Urquhart J.,   White G.~J.,  2008, Astronomy \& Astrophysics, 477, 557

\bibitem[\protect\citeauthoryear{Okada et~al.,}{Okada et~al.}{2024}]{okada2024bright}
Okada Y.,  et~al., 2024, Astronomy \& Astrophysics, 690, A45

\bibitem[\protect\citeauthoryear{Panagia}{Panagia}{1973}]{panagia1973some}
Panagia N.,  1973, Astronomical Journal, vol. 78, p. 929-934 (1973)., 78, 929

\bibitem[\protect\citeauthoryear{Patel, Goldsmith, Snell, Hezel  \& Xie}{Patel et~al.}{1995}]{patel1995large}
Patel N.~A.,  Goldsmith P.~F.,  Snell R.~L.,  Hezel T.,   Xie T.,  1995, Astrophysical Journal v. 447, p. 721, 447, 721

\bibitem[\protect\citeauthoryear{Pineda, Caselli  \& Goodman}{Pineda et~al.}{2008}]{pineda2008co}
Pineda J.~E.,  Caselli P.,   Goodman A.~A.,  2008, The Astrophysical Journal, 679, 481

\bibitem[\protect\citeauthoryear{Porel, Soam, Karoly, Chung  \& Lee}{Porel et~al.}{2025a}]{porel2025investigating}
Porel P.,  Soam A.,  Karoly J.,  Chung E.~J.,   Lee C.~W.,  2025a, Monthly Notices of the Royal Astronomical Society, 542, 2953

\bibitem[\protect\citeauthoryear{Porel, Soam, Karoly, Chung, Lee, Kim, Gupta  \& Sharma}{Porel et~al.}{2025b}]{porel2025unbound}
Porel P.,  Soam A.,  Karoly J.,  Chung E.~J.,  Lee C.~W.,  Kim S.,  Gupta S.,   Sharma N.,  2025b, The Astrophysical Journal, 995, 29

\bibitem[\protect\citeauthoryear{Puschnig}{Puschnig}{2020}]{puschnig2020dense}
Puschnig J.,  2020, Zenodo

\bibitem[\protect\citeauthoryear{Ramesh}{Ramesh}{1995}]{ramesh1995study}
Ramesh B.,  1995, Monthly Notices of the Royal Astronomical Society, 276, 923

\bibitem[\protect\citeauthoryear{Rapson, Pipher, Gutermuth, Megeath, Allen, Myers  \& Allen}{Rapson et~al.}{2014}]{rapson2014spitzer}
Rapson V.~A.,  Pipher J.~L.,  Gutermuth R.~A.,  Megeath S.~T.,  Allen T.~S.,  Myers P.~C.,   Allen L.~E.,  2014, The Astrophysical Journal, 794, 124

\bibitem[\protect\citeauthoryear{Rawat et~al.,}{Rawat et~al.}{2023}]{rawat2023probing}
Rawat V.,  et~al., 2023, Monthly Notices of the Royal Astronomical Society, 521, 2786

\bibitem[\protect\citeauthoryear{Rawat et~al.,}{Rawat et~al.}{2024}]{rawat2024giant}
Rawat V.,  et~al., 2024, Monthly Notices of the Royal Astronomical Society, 528, 2199

\bibitem[\protect\citeauthoryear{Reipurth}{Reipurth}{1983}]{reipurth1983star}
Reipurth B.,  1983, Astronomy and Astrophysics, vol. 117, no. 2, Jan. 1983, p. 183-198., 117, 183

\bibitem[\protect\citeauthoryear{Reipurth, Rodr{\'\i}guez, Anglada  \& Bally}{Reipurth et~al.}{2002}]{reipurth2002radio}
Reipurth B.,  Rodr{\'\i}guez L.~F.,  Anglada G.,   Bally J.,  2002, The Astronomical Journal, 124, 1045

\bibitem[\protect\citeauthoryear{Reipurth, Yu, Moriarty-Schieven, Bally, Aspin  \& Heathcote}{Reipurth et~al.}{2004}]{reipurth2004deep}
Reipurth B.,  Yu K.~C.,  Moriarty-Schieven G.,  Bally J.,  Aspin C.,   Heathcote S.,  2004, The Astronomical Journal, 127, 1069

\bibitem[\protect\citeauthoryear{Saha, Maheswar, Ojha, Baug  \& Neha}{Saha et~al.}{2022}]{saha2022investigation}
Saha P.,  Maheswar G.,  Ojha D.,  Baug T.,   Neha S.,  2022, Monthly Notices of the Royal Astronomical Society: Letters, 515, L67

\bibitem[\protect\citeauthoryear{Spitzer~Jr}{Spitzer~Jr}{2008}]{spitzer2008physical}
Spitzer~Jr L.,  2008, Physical processes in the interstellar medium.
John Wiley \& Sons

\bibitem[\protect\citeauthoryear{Stahler \& Palla}{Stahler \& Palla}{2008}]{stahler2008formation}
Stahler S.~W.,  Palla F.,  2008, The formation of stars.
John Wiley \& Sons

\bibitem[\protect\citeauthoryear{Stobie}{Stobie}{1980}]{stobie1980application}
Stobie R.,  1980, in Applications of Digital Image Processing to Astronomy. pp 208--212

\bibitem[\protect\citeauthoryear{Sugitani, Fukui  \& Ogura}{Sugitani et~al.}{1991}]{sugitani1991catalog}
Sugitani K.,  Fukui Y.,   Ogura K.,  1991, Astrophysical Journal Supplement Series (ISSN 0067-0049), vol. 77, Sept. 1991, p. 59-66., 77, 59

\bibitem[\protect\citeauthoryear{Sung \& Bessell}{Sung \& Bessell}{2010}]{sung2010initial}
Sung H.,  Bessell M.~S.,  2010, The Astronomical Journal, 140, 2070

\bibitem[\protect\citeauthoryear{Turner}{Turner}{1976}]{turner1976value}
Turner D.~G.,  1976, Astrophysical Journal, vol. 210, Nov. 15, 1976, pt. 1, p. 65-75. Research supported by the National Research Council of Canada., 210, 65

\bibitem[\protect\citeauthoryear{Vallenari et~al.,}{Vallenari et~al.}{2023}]{vallenari2023gaia}
Vallenari A.,  et~al., 2023, Astronomy \& Astrophysics, 674, A1

\bibitem[\protect\citeauthoryear{Virtanen et~al.,}{Virtanen et~al.}{2020}]{virtanen2020scipy}
Virtanen P.,  et~al., 2020, Nature methods, 17, 261

\bibitem[\protect\citeauthoryear{Walsh, Ogura  \& Reipurth}{Walsh et~al.}{1992}]{walsh1992two}
Walsh J.,  Ogura K.,   Reipurth B.,  1992, Monthly Notices of the Royal Astronomical Society, 257, 110

\bibitem[\protect\citeauthoryear{Wolf-Chase, Moriarty-Schieven, Fich  \& Barsony}{Wolf-Chase et~al.}{2003}]{wolf2003star}
Wolf-Chase G.,  Moriarty-Schieven G.,  Fich M.,   Barsony M.,  2003, Monthly Notices of the Royal Astronomical Society, 344, 809

\bibitem[\protect\citeauthoryear{Zhou et~al.,}{Zhou et~al.}{2023}]{zhou2023high}
Zhou J.,  et~al., 2023, Astronomy \& Astrophysics, 676, A69

\bibitem[\protect\citeauthoryear{Zhou, Wyrowski, Neupane, Christensen, Menten, Li  \& Liu}{Zhou et~al.}{2024}]{zhou2024high}
Zhou J.,  Wyrowski F.,  Neupane S.,  Christensen I.~B.,  Menten K.,  Li S.,   Liu T.,  2024, Astronomy \& Astrophysics, 682, A128

\bibitem[\protect\citeauthoryear{Zhou, Li  \& Chen}{Zhou et~al.}{2025}]{zhou2025kinematics}
Zhou J.-X.,  Li G.-X.,   Chen B.-Q.,  2025, Monthly Notices of the Royal Astronomical Society, 542, 52

\makeatother
\end{thebibliography}








\bsp	
\label{lastpage}
\end{document}